\documentclass[preprintnumbers,nofootinbib,
superscriptaddress,amsmath,floatfix,prd]{revtex4}

\usepackage{amssymb}  % \gtrsim, \geqslant, etc: see amsguide.ps
\usepackage{graphicx,subfigure}
\usepackage{exscale}
\usepackage{textcomp}
\usepackage{enumerate}
\usepackage [english]{babel}
\usepackage [autostyle, english = american]{csquotes}
\MakeOuterQuote{"}
\usepackage{color}

\usepackage[normalem]{ulem}  % \sout{old text} for strikeout

\renewcommand\l{\lambda}

\renewcommand\u{\upsilon}

\newcommand{\non}{\nonumber\\}

\newcommand{\be}{\begin{equation}}
\newcommand{\ee}{\end{equation}}
\newcommand{\bea}{\begin{eqnarray}}
\newcommand{\eea}{\end{eqnarray}}
\newcommand{\ba}[1]{\begin{array}{#1}}
\newcommand{\ea}{\end{array}}

\newcommand{\Tr}{{\rm Tr}}

\newcommand{\bm}[1]{\mbox{\boldmath${#1}$}}

\begin{document}

\title{New color-magnetic defects in dense quark matter}

\author{Alexander Haber}
\email{ahaber@hep.itp.tuwien.ac.at}
\affiliation{Institut f\"{u}r Theoretische Physik, Technische Universit\"{a}t Wien, 1040 Vienna, Austria}
\affiliation{Mathematical Sciences and STAG Research Centre, University of Southampton, Southampton SO17 1BJ, United Kingdom}

\author{Andreas Schmitt}
\email{a.schmitt@soton.ac.uk}
\affiliation{Mathematical Sciences and STAG Research Centre, University of Southampton, Southampton SO17 1BJ, United Kingdom}

\date{6 August 2020}

\begin{abstract}

Color-flavor locked (CFL) quark matter expels color-magnetic fields due to the Meissner effect. One of these fields carries an admixture of the ordinary abelian magnetic field and therefore flux tubes may 
form if CFL matter is exposed to a magnetic field, possibly in the interior of neutron stars or in quark stars. We employ a Ginzburg-Landau approach for three massless quark flavors, 
which takes into account the multi-component nature of color superconductivity. Based on the weak-coupling 
expressions for the Ginzburg-Landau parameters, we identify the regime where CFL is a type-II color superconductor and compute the radial profiles of different color-magnetic flux tubes. Among the configurations without baryon circulation we find a new solution that is energetically preferred over the flux tubes previously 
discussed in the literature in the parameter regime relevant for compact stars.  
Within the same setup, we also find a new defect in the 2SC phase, namely magnetic domain walls, 
which emerge naturally from the previously studied flux tubes if a more general ansatz for the 
order parameter is used. Color-magnetic defects in the interior of compact stars allow for sustained deformations of the star, potentially strong enough to produce detectable gravitational waves.
 
\end{abstract}

\maketitle

%\tableofcontents

%%%%%%%%%%%%%%%%%%%%%%%%%%%%%%%%%%%%%%%%%%%%%%%%%%%%%
\section{Introduction}
%%%%%%%%%%%%%%%%%%%%%%%%%%%%%%%%%%%%%%%%%%%%%%%%%%%%%

Ordinary superconductivity can be destroyed by an external magnetic field: either partially, by the formation of magnetic flux tubes if the superconductor is of type II, or completely,
if the external field is sufficiently large \cite{abrikosov1957,tinkham2004introduction,RevModPhys.82.109}. Here we investigate the fate of {\it color} superconductivity in three-flavor 
quark matter in the presence of an ordinary external magnetic field, with an emphasis on the magnetic defects created in type-II color superconductors. 

At the highest densities, three-flavor quark matter is in the color-flavor locked (CFL) phase \cite{Alford:1997zt,Alford:1998mk,Alford:2007xm}, where all quarks participate in Cooper 
pairing. All Cooper pairs are neutral with respect to a certain combination of the electromagnetic gauge field and the eighth gluon gauge field. The corresponding magnetic field,
which we call $\tilde{B}$, can penetrate the CFL phase, while the magnetic field corresponding to the orthogonal combination, termed $\tilde{B}_8$, and the fields corresponding 
to the other seven gluons are expelled due to the Meissner effect. Since an ordinary magnetic field $B$ has a $\tilde{B}_8$ component it will eventually destroy the CFL phase and, in the type-II regime for intermediate field strengths, will lead to the formation of magnetic flux tubes that carry $\tilde{B}_8$ flux. 

%%%%%%%%%%%%%%%%%%%%%%%%%%%%%%%%%%%%%%%%%%%%%%%%%%%%%
\subsection{Method and main ideas}
%%%%%%%%%%%%%%%%%%%%%%%%%%%%%%%%%%%%%%%%%%%%%%%%%%%%%

Magnetic flux tubes in CFL have been studied in Refs.\ \cite{Iida:2002ev,Iida:2004if} within a Ginzburg-Landau 
approach \cite{Bailin:1983bm,Iida:2000ha,Iida:2001pg}, including an analysis of whether CFL is a type-I or type-II superconductor. 
This question was also
addressed within the same approach in Ref.\ \cite{Giannakis:2003am}, by calculating the surface energy. In these works, CFL was effectively described as a 
two-component superconductor, where the two components have different charges with respect to the rotated color gauge field $\tilde{A}_8^\mu$. 
In this paper, we employ the same Ginzburg-Landau approach, but improve the known results in several ways. Firstly, we make use of 
recently gained understanding about two-component superconductivity, in particular the unconventional behavior of such systems in the type-I/type-II transition region, which has been discussed for instance in the context of dense nuclear matter  \cite{Haber:2017kth} 
and two-band superconductivity \cite{PhysRevB.72.180502,PhysRevLett.105.067003}. Secondly, we show that the 
CFL phase is, upon increasing the magnetic field, superseded by the so-called 2SC phase (except for very small values of the strong coupling constant), which is indicative of the kind of flux tubes that develop in CFL. We show, thirdly, that a new kind 
of flux tubes is energetically preferred in the parameter regime that is relevant for applications to compact stars. This new flux tube configuration is found by allowing all three diagonal components of the order parameter to be different, in contrast to the two-component approach in the 
literature. The total winding of the three components is minimized by setting the winding number of one component to zero, 
resulting in a CFL flux tube with a 2SC-like core. By computing the critical magnetic field at which flux tubes start to populate the system, 
we shall demonstrate that this configuration is favored over the previously discussed CFL flux tubes with an unpaired core. 

We will also study flux tubes in 2SC itself. Since the 2SC phase is a single-component superconductor, the flux tube configuration considered in Ref.\ \cite{Alford:2010qf} appears to be unique, analogous to ordinary superconductors. However, our general setup allows us to check whether additional color-flavor components of the order parameter are induced in the core of a 2SC flux tube. We find that this is indeed the case. These new flux tube solutions can reduce their energy by increasing their winding number and thus their radius, eventually resulting in a domain wall rather than a one-dimensional string. 

By using the purely bosonic Ginzburg-Landau theory we neglect any effect of the charges of the constituents of the Cooper pairs, and a fermionic approach would have to be used to go beyond this 
approximation \cite{Ferrer:2005vd,Ferrer:2006vw,Noronha:2007wg,Fukushima:2007fc}. Moreover, as usual, the  
Ginzburg-Landau approach is strictly speaking only valid for small condensates, for instance for temperatures close to the critical temperature. Apart from these restrictions, our results are very general since they do not depend on the underlying microscopic theory. For our main numerical results, however, we do not 
investigate the complete parameter space of the Ginzburg-Landau potential, but rather restrict ourselves  to  the 
weak-coupling form of the parameters and extrapolate the results to larger values of the coupling, which are expected in an astrophysical environment. We also work in the simplified scenario of vanishing quark masses, and 
it remains to be seen how our results are modified if the strange quark mass is taken into account; mass terms were included in the Ginzburg-Landau approach in 
Refs.\ \cite{Iida:2003cc,Iida:2004cj,Schmitt:2010pf}. 

%%%%%%%%%%%%%%%%%%%%%%%%%%%%%%%%%%%%%%%%%%%%%%%%%%%%%
\subsection{Relation to superfluid vortices in CFL}
%%%%%%%%%%%%%%%%%%%%%%%%%%%%%%%%%%%%%%%%%%%%%%%%%%%%%

All flux tubes we discuss in detail have a vanishing baryon circulation far away from the flux tube. 
In other words, the flux tubes we are interested in can only be induced by a magnetic field, not by rotation. Flux tubes 
that do have baryon circulation, in particular the so-called semi-superfluid vortices, have been discussed extensively in the literature, 
for instance in Refs.\ \cite{Balachandran:2005ev,Eto:2009kg,Vinci:2012mc,Alford:2016dco}, for a review see Ref.\ \cite{Eto:2013hoa}.
These vortices, just like the vortices in an ordinary superfluid, have a logarithmically divergent energy, and a finite system or a lattice of 
vortices is required to regularize this divergence. The flux tubes we discuss here, just like the flux tubes in an ordinary 
superconductor, do not show this divergence and their energy is finite even in an infinite volume. To put our discussion into a wider context, we shall briefly discuss how all line defects, with and without baryon circulation, with and without color-magnetic flux,  are  obtained by choosing different triples of winding numbers of the three order parameter components. 

In contrast to the CFL vortices, the flux tubes discussed here are not protected by 
topology \cite{Eto:2013hoa}. This means that configurations with different windings are continuously connected. In particular, the configurations we consider are continuously connected to the zero-winding configuration (not unlike the so-called "semilocal cosmic strings" \cite{Vachaspati:1991dz}), i.e., they can be unwound into "nothing" 
without encountering a discontinuity.
Since such a discontinuity typically translates into an energy barrier, one might 
question the stability of the objects we consider in this paper. However, the main result of our calculation is a critical magnetic field at which 
the flux tube is energetically preferred over the configuration without flux tube. Therefore, even though we do not explicitly prove  local stability by introducing fluctuations about the flux tube state, the magnetic field stabilizes the flux tube and by comparing free energies 
we establish  global stability. (Our ansatz is not completely general in color-flavor space, i.e., while we will prove that the flux tube cannot 
decay into "nothing" at a sufficiently large magnetic field, we can, strictly speaking, not exclude that it decays into more exotic 
color-magnetic flux tubes.) 

%%%%%%%%%%%%%%%%%%%%%%%%%%%%%%%%%%%%%%%%%%%%%%%%%%%%%
\subsection{Astrophysical implications}
%%%%%%%%%%%%%%%%%%%%%%%%%%%%%%%%%%%%%%%%%%%%%%%%%%%%%

Color-magnetic defects in CFL and 2SC quark matter are very interesting for the phenomenology of quark stars or neutron stars
with a quark matter core. The critical magnetic fields we compute here -- as already suggested from previous work -- are most likely too large to be reached in compact stars. Nevertheless, there might be other mechanisms to create magnetic defects in quark matter. As argued in Ref.\ \cite{Alford:2010qf}, flux tubes can form if quark matter is cooled into a color-superconducting phase at a given, approximately constant magnetic field. It is then a dynamic question how and on which time 
scale the magnetic field is expelled from the system. A full dynamical simulation of the expulsion of the magnetic field is extremely 
complicated and most likely involves the formation of flux tubes or domain walls, see for instance Ref.\ \cite{1991PhRvL..66.3071L} for such a study the context of ordinary superconductors. While our results only concern equilibrium configurations they 
show, to the very least, that new defects, so far overlooked in the literature, should be taken into account in this discussion.  

It has been argued that the color-flux tubes thus created
support a deformation of the rotating star ("color-magnetic mountains"). This deformation gives rise to a continuous emission of gravitational waves because of the misalignment of rotational and magnetic axes \cite{Glampedakis:2012qp}.
(A different mechanism in quark matter to support continuous gravitational waves is the formation of a crystalline phase  \cite{Lin:2007rz,Haskell:2007sh,Knippel:2009st,Anglani:2013gfu}.)
The larger energy (and the only slightly smaller number) of the color-magnetic flux tubes compared to  flux tubes in superconducting nuclear matter makes this mechanism particularly efficient and the resulting gravitational waves potentially detectable. Our calculation
provides a quantitative, numerical  calculation of the flux tube energy, putting the estimates used in Ref.\ \cite{Glampedakis:2012qp}
on solid ground. It also slightly changes this estimate due to the new flux tube configuration, although this change is small 
compared to the uncertainties involved in the estimate of the ellipticity of the star. 

%%%%%%%%%%%%%%%%%%%%%%%%%%%%%%%%%%%%%%%%%%%%%%%%%%%%%
\subsection{Structure of the paper}
%%%%%%%%%%%%%%%%%%%%%%%%%%%%%%%%%%%%%%%%%%%%%%%%%%%%%
Our paper is organized as follows. In Sec.\ \ref{sec:GL} we introduce the Ginzburg-Landau potential and our ansatz for the order parameter. Then,  as a necessary preparation for the study of the flux tubes, in Sec.\ \ref{sec:hom} we discuss the homogeneous phases and the phase diagram at nonzero external magnetic field. We turn to the CFL flux tubes in Sec.\ \ref{sec:CFL}, with a classification 
of the flux tubes and their radial profiles shown in Sec.\ \ref{sec:circ}. In Sec.\ \ref{sec:2SC} we discuss 2SC flux tubes and domain 
walls and present the corresponding profiles in Sec.\ \ref{sec:profiles}. The main results, putting together the phase diagram of the homogeneous phases with the critical fields for the magnetic defects, are discussed in Sec.\ \ref{sec:results}, and we give a brief summary and outlook, including astrophysical implications, in Sec.\ \ref{sec:summary}.
Our convention for the metric tensor is $g^{\mu\nu}={\rm diag}(1,-1,-1,-1)$. We work in natural units $\hbar=c=k_B=1$ and use Heaviside-Lorentz units for the gauge fields, in which the 
elementary charge is $e=\sqrt{4\pi\alpha}\simeq 0.3$. These are the units used in the most closely related literature about the 
CFL phase, for instance Ref.\ \cite{Giannakis:2003am}. Note, however, that Gaussian units are used in other literature on multi-component superconductors, for instance in Ref.\ \cite{Haber:2017kth}.

%%%%%%%%%%%%%%%%%%%%%%%%%%%%%%%%%%%%%%%%%%%%%%%%%%%%%%%%%%
\section{Setup}
\label{sec:GL}
%%%%%%%%%%%%%%%%%%%%%%%%%%%%%%%%%%%%%%%%%%%%%%%%%%%%%%%%%%

%%%%%%%%%%%%%%%%%%%%%%%%%%%%%%%%%%%%%%%%%%%%%%%%%%%%%%%%%%
\subsection{General form of Ginzburg-Landau potential}
%%%%%%%%%%%%%%%%%%%%%%%%%%%%%%%%%%%%%%%%%%%%%%%%%%%%%%%%%%

The order parameter $\Psi$ for spin-zero Cooper pairing of three-flavor, three-color quark matter is an anti-triplet  
in color and flavor space, $\Psi \in [\bar{3}]_c\otimes [\bar{3}]_f$. Both anti-triplets are spanned by three anti-symmetric 
$3\times 3$ matrices, say $(J_i)_{jk} = -i\epsilon_{ijk}$ in color space and $(I_i)_{jk} = -i\epsilon_{ijk}$ in flavor space. 
We can thus introduce the components $\Phi_{ij}$ of the order parameter in the given basis via 
\be
\Psi =  \Phi_{ij} J_i\otimes I_j \, . 
\ee
Later, we shall only work with the $3\times 3$ matrix $\Phi$, not with the $9\times 9$ tensor $\Psi$, and simply refer to $\Phi$ as the order parameter. The structure of $\Phi$ determines the pairing pattern, i.e., the particular color-superconducting phase. For example, 
$\Phi_{ij} \propto \delta_{ij}$ for CFL and $\Phi_{ij} \propto \delta_{i3}\delta_{j3}$ for 2SC. In general, there are two order parameters $\Psi_L$ and $\Psi_R$ for pairing in the left-handed and right-handed sectors. They are different for instance if kaon condensation is considered \cite{Schmitt:2010pf}. Here we assume $\Psi_L=\Psi_R\equiv \Psi$.
The Ginzburg-Landau potential up to quartic order in $\Psi$ is \cite{Giannakis:2003am}
\bea \label{U}
U &=& -3\Big\{\Tr[(D_0\Psi)^\dag(D_0\Psi)]-u^2\Tr[(D_i\Psi)^\dag(D_i\Psi)]\Big\}+k\Tr[\Psi^\dag\Psi]+\frac{l_1}{2}\Tr[(\Psi^\dag\Psi)^2]+\frac{l_2}{2}(\Tr[\Psi^\dag\Psi])^2 \non[2ex]
&&+\frac{1}{4}F_{\mu\nu}^aF_a^{\mu\nu}+\frac{1}{4}F_{\mu\nu}F^{\mu\nu} \, , 
\eea
where $u^2 = \frac{1}{3}$, where  $F_{\mu\nu}^a=\partial_\mu A_\nu^a-\partial_\nu A_\mu^a+gf^{abc}A_\mu^bA_\nu^c$ are the gluonic field strength tensors  with $a=1,\ldots,8$,  the color gauge fields $A_\mu^a$, the strong coupling constant $g$,  and the SU(3) structure constants $f^{abc}$, and  
$F_{\mu\nu}=\partial_\mu A_\nu-\partial_\nu A_\mu$ is the electromagnetic field strength tensor with the electromagnetic gauge field $A_\mu$. The parameters $k$, $l_1$, $l_2$ 
can be computed in the weak-coupling limit from perturbation theory \cite{Iida:2000ha}. The covariant derivative is 
\bea\label{DPsi}
D_\mu \Psi &=& \partial_\mu \Psi +igA_\mu^a\Phi_{ij}(T_a J_i+J_i T_a^T)\otimes I_j +ie A_\mu \Phi_{ij} J_i\otimes (QI_j+I_j Q^T) 
\, ,
\eea
where $T_a=\lambda_a/2$, with the Gell-Mann matrices $\lambda_a$,  such that $\Tr[T^aT^b]=\frac{1}{2}\delta^{ab}$,  where $e$ is the elementary electric charge, and where $Q={\rm diag}(q_1,q_2,q_3)$ is the 
$U(1)$ charge generator in flavor space with the individual electric charges of the quarks $q_1$, $q_2$, $q_3$.

For simplicity, we shall work in the massless limit throughout the paper, such that flavor symmetry is only 
broken by the electric charges, not by the quark masses. In particular, there is no distinction between $d$ and $s$ quarks in our approximation. We can write the covariant derivative as  
\bea
D_\mu \Psi 
&=& (D_\mu\Phi)_{ij} J_i\otimes I_j\, ,
\eea
with
\be \label{Dphi}
D_\mu\Phi = \partial_\mu\Phi -igA_\mu^a T_a^T\Phi+ieA_\mu \Phi \bar{Q} \, ,
\ee
where we have used $T_a J_i+J_i T_a^T = -(T_a)_{ij}J_j$ and  $QI_j+I_j Q^T = \bar{Q}_{jk} I_k$ with 
$\bar{Q}= {\rm diag}(q_2+q_3,q_1+q_3,q_1+q_2)$. Since the electric charges of $u$, $d$ and $s$ quarks add up to zero, we have 
$\bar{Q} = -Q$, and thus it is not strictly necessary to  introduce the notation $\bar{Q}$. But, 
one should keep in mind that the relevant charge matrix contains the 
charges of Cooper pairs, not of individual quarks, as the notation $-Q$ instead of $\bar{Q}$ would have suggested.  
We can now perform the trace over the 9-dimensional color-flavor space in Eq.\ (\ref{U}) and write the Ginzburg-Landau potential in terms of $\Phi$, 
\bea \label{UPhi}
U &=& -12\Big\{\Tr[(D_0\Phi)^\dag(D_0\Phi)]-u^2\Tr[(D_i\Phi)^\dag(D_i\Phi)]\Big\}+4k \Tr[\Phi^\dag\Phi]+l_1\Tr[(\Phi^\dag\Phi)^2]+
(l_1+8l_2) (\Tr[\Phi^\dag\Phi])^2 \non[2ex]
&&+\frac{1}{4}F_{\mu\nu}^aF_a^{\mu\nu}+\frac{1}{4}F_{\mu\nu}F^{\mu\nu}
 \, , 
\eea
where now the traces are taken over the 3-dimensional order parameter space.

%%%%%%%%%%%%%%%%%%%%%%%%%%%%%%%%%%%%%%%%%%%%%%%%%%%%%%%%%%
\subsection{Superfluid velocity}
%%%%%%%%%%%%%%%%%%%%%%%%%%%%%%%%%%%%%%%%%%%%%%%%%%%%%%%%%%

Superfluid vortices are characterized by a nonzero circulation around the vortex. We shall see that line defects in CFL 
can carry magnetic flux {\it and} baryon circulation. Therefore, we first derive a general expression for the superfluid velocity, which 
can then be used to compute the baryon circulation for particular flux tube solutions, see Sec.\ \ref{sec:circ}. The superfluid velocity is computed in analogy to the case of a scalar field \cite{2013PhRvD..87f5001A}; for a derivation in the context of CFL see 
Refs.\ \cite{Iida:2001pg,Eto:2013hoa}.
We first introduce an overall phase $\psi$ associated with 
baryon number conservation $U(1)_B$,
\be
\Phi = e^{i\psi}\Delta \, .
\ee
This allows us to compute the baryon four-current via 
\bea
j^\mu = -\frac{\partial U}{\partial (\partial_\mu\psi)} \, . 
\eea
We find 
\bea
j^0&=& 12 i\Tr[(D^0\Phi)^\dag\Phi-\Phi^\dag(D^0\Phi)] \, , \qquad j^i= 12 u^2 i\Tr[(D^i\Phi)^\dag\Phi-\Phi^\dag(D^i\Phi)] \, .
\eea
The superfluid four-velocity $v^\mu$ is defined through $j^\mu = n_s v^\mu$ 
with the superfluid density $n_s$ and $v_\mu v^\mu=1$. 
With $v^\mu=\gamma(1,{\bf v}_s)$, where $\gamma$ is the usual Lorentz factor, the components of the superfluid three-velocity 
${\bf v}_s$ become
\be \label{superv}
({\bf{v}_s})_i = \frac{j^i}{j^0} = \frac{u^2}{4\mu_q}\frac{i\Tr[(D^i\Phi)^\dag\Phi-\Phi^\dag(D^i\Phi)]}{\Tr[\Phi^\dag\Phi]} \, , 
\ee
where we have assumed $\Delta$ to be time-independent, set the temporal components of the gauge fields to zero, \mbox{$A_0=A^a_0=0$}, 
and introduced the quark chemical potential $\mu_q$ through the time dependence of the phase, $\partial_0 \psi= 2\mu_q$, where 
the factor 2 arises from the diquark nature of the order parameter.

%%%%%%%%%%%%%%%%%%%%%%%%%%%%%%%%%%%%%%%%%%%%%%%%%%%%%%%%%%
\subsection{Ansatz and Gibbs free energy}
%%%%%%%%%%%%%%%%%%%%%%%%%%%%%%%%%%%%%%%%%%%%%%%%%%%%%%%%%%

We evaluate the potential (\ref{UPhi}) for the diagonal order parameter $\Phi = \frac{1}{2}{\rm diag}(\phi_1,\phi_2,\phi_3)$, with the complex 
scalar fields $\phi_1$, $\phi_2$, $\phi_3$. Allowing all three diagonal components to be different is a more general ansatz than used in the literature before. It is not the most general ansatz because the reduced symmetry due to the electric charges of the quarks
(and quark masses if they were taken into account) does not allow to rotate an arbitrary order parameter matrix into an equivalent diagonal form. 
For our diagonal order parameter, it is consistent with the non-abelian Maxwell equations to set all gauge fields corresponding to the non-diagonal 
$SU(3)$ generators to zero, $A_1^\mu=A_2^\mu=A_4^\mu=A_5^\mu=A_6^\mu=A_7^\mu=0$. 
The eighth gluon and the photon mix, which can for instance be seen in a microscopic calculation of the gauge boson polarization tensor \cite{Schmitt:2003aa}. This mixing can also be derived within the Ginzburg-Landau approach by computing the 
magnetic fields in the CFL phase in the presence of an externally applied magnetic field, which we will do in Sec.\ \ref{sec:hom}. 
 We anticipate this mixing by defining the rotated gauge fields 
\begin{subequations} \label{mix0}
\bea
\tilde{A}_\mu^8 &=& \cos\theta \,A_\mu^8+\sin\theta\, A_\mu \, , \\[2ex]
\tilde{A}_\mu &=& -\sin\theta \,A_\mu^8+\cos\theta\, A_\mu \, , 
\eea
\end{subequations}
with the mixing angle given by
\be \label{costheta}
\cos\theta = \frac{\sqrt{3}g}{\sqrt{3g^2+4e^2}} \, ,  \qquad \sin\theta = - \frac{2e}{\sqrt{3g^2+4e^2}} \, .
\ee
In the new rotated basis, the magnetic field $\tilde{B}_8$ experiences a Meissner effect in the CFL phase and the magnetic field 
$\tilde{B}$ penetrates the CFL phase unperturbed, if the quark flavors in the charge matrix are ordered $(d,s,u)$, such that $Q={\rm diag}(-1/3,-1/3,2/3)$ is proportional to $T_8$. If the order $(u,d,s)$ is used, the mixing between the gauge fields involves $A_\mu^3$ \cite{Iida:2001pg}. We shall work with the more convenient order $(d,s,u)$ in the CFL phase, but change to $(u,d,s)$ in Sec.\ \ref{sec:2SC}, where 
we discuss magnetic defects in the 2SC phase. 

We set all electric fields to zero, and only keep 
the magnetic fields ${\bf B}_3 = \nabla\times{\bf A}_3$, $\tilde{\bf B}_8 = \nabla\times\tilde{\bf A}_8$, and $\tilde{\bf B} = \nabla\times\tilde{\bf A}$.  We also ignore all time dependence since we are only interested in equilibrium configurations. 
Putting all this together yields 
the potential 
\be \label{UUB}
U = U_0 + \frac{\tilde{\bf B}^2}{2} \, , 
\ee
with 
\bea \label{U123}
U_0 &=& \frac{{\bf B}_3^2}{2}+ \frac{\tilde{\bf B}_8^2}{2} +\left|\left(\nabla+i\frac{g}{2}{\bf A}_3+i\tilde{g}_8\tilde{\bf A}_8\right)\phi_1\right|^2
+\left|\left(\nabla-i\frac{g}{2}{\bf A}_3+i\tilde{g}_8\tilde{\bf A}_8\right)\phi_2\right|^2+\left|\left(\nabla-2i\tilde{g}_8\tilde{\bf A}_8\right)\phi_3\right|^2 \non[2ex]
&&-\mu^2(|\phi_1|^2+|\phi_2|^2+|\phi_3|^2)+\lambda(|\phi_1|^4+|\phi_2|^4+|\phi_3|^4)-2h(|\phi_1|^2|\phi_2|^2+|\phi_1|^2|\phi_3|^2+|\phi_2|^2|\phi_3|^2) \, . 
\eea
We have separated the rotated field $\tilde{\bf B}$ because all scalar fields are neutral with respect to the corresponding charge, and the only contribution
is the trivial $\tilde{\bf B}^2$ term. We have denoted the coupling to the rotated color field $\tilde{\bf A}_8$ by
\be \label{g8}
\tilde{g}_8 \equiv \frac{g}{2\sqrt{3}\cos\theta}  \, ,
\ee
and introduced the new Ginzburg-Landau parameters 
\begin{subequations} \label{weak123}
\bea
\mu^2 &=& -k \simeq \frac{48\pi^2}{7\zeta(3)}T_c(T_c-T) \, , \\[2ex]
\lambda&=&\frac{l_1}{8}+\frac{l_2}{2}  \simeq\frac{72\pi^4}{7\zeta(3)}\frac{T_c^2}{\mu_q^2}  \, ,\\[2ex]
h&=&-\left(\frac{l_1}{16}+\frac{l_2}{2}\right)  \simeq-\frac{36\pi^4}{7\zeta(3)}\frac{T_c^2}{\mu_q^2} \, .
\eea
\end{subequations} 
In the last expression of each line, the weak-coupling results have been used\footnote{We are using the convention of Ref.\ \cite{Giannakis:2003am}. To compare with 
Refs.\ \cite{Iida:2002ev,Iida:2004if,Eto:2013hoa}, the order parameter has to be rescaled as 
\be
\Phi \to \sqrt{\frac{3}{7\zeta(3)}}\frac{\pi^2T_c}{2\mu_q} \Phi\, .\nonumber 
\ee
}
  with the temperature $T$
and the critical temperature for color superconductivity $T_c$. 
(At weak coupling, although the relation between the 
critical temperature and the zero-temperature gap differs from phase to phase \cite{Schmitt:2002sc}, the absolute values of the critical temperatures of CFL and 2SC are the same.) The potential (\ref{U123}) describes three massless bosonic fields which have 
the same chemical potential $\mu$, the same self-interaction given by $\lambda$, interact pairwise with the same coupling constant $h$, and have different charges with respect to the three gauge fields. (In comparison, the model in Ref.\ \cite{Haber:2017kth} contains two massive scalar fields with different chemical potentials and different self-couplings, including derivative coupling terms
between the fields.)
For $\phi_1=\phi_2$ the system is
 neutral  with respect to $A^\mu_3$ at every point in space and we recover the potential used in Ref.\  \cite{Giannakis:2003am}. Since we allow for $\phi_1\neq \phi_2$, we must keep 
${\bf A}_3$.

We are interested in the phase structure in an externally given homogeneous magnetic field ${\bf H}$, which, without 
loss of generality, we align with the $z$-direction, ${\bf H}=H{\bf e}_z$ with $H\ge 0$. Therefore, 
we need to consider the Gibbs free energy 
\be
G = \int d^3{\bf r}\,(U-{\bf H}\cdot{\bf B})  = \int d^3{\bf r} \left[U_0+\frac{\tilde{\bf B}^2}{2}-{\bf H}\cdot(\tilde{\bf B}\cos\theta+\tilde{\bf B}_8\sin\theta)\right] \, .
\ee
Since $\tilde{\bf A}$ does not couple to the three condensates, its equation of motion is trivially fulfilled by any constant $\tilde{\bf B}$ and 
the Gibbs free energy is minimized by $\tilde{\bf B}=\tilde{B}{\bf e}_z$ with
\be
\tilde{B} = H\cos\theta \,, 
\ee
such that we can write the Gibbs free energy density as 
\be \label{GV}
\frac{G}{V} = -\frac{H^2\cos^2\theta}{2}+ \frac{1}{V}\int d^3{\bf r} \left(U_0-{\bf H}\cdot\tilde{\bf B}_8\sin\theta\right) \, ,
\ee
 where $V$ is the total volume of the system.

%%%%%%%%%%%%%%%%%%%%%%%%%%%%%%%%%%%%%%%%%%%%%%%%%%%%%%%%%%
\subsection{Strategy of our calculation}
%%%%%%%%%%%%%%%%%%%%%%%%%%%%%%%%%%%%%%%%%%%%%%%%%%%%%%%%%%

In order to identify the region in parameter space where magnetic flux tubes form, we need to compute the three critical magnetic fields $H_c$, $H_{c1}$ and $H_{c2}$. The critical field $H_c$ follows from a simple comparison of Gibbs free energies of the homogeneous phases. The critical field $H_{c2}$ 
is defined as the maximal magnetic field that can be sustained in the superconducting phase, assuming a second order phase transition 
from the flux tube phase to the normal phase. Also the calculation of $H_{c2}$ is simple because the equations of motion can be linearized in the condensate. Only
$H_{c1}$, the field at which a single flux tube enters the superconductor, requires a fully numerical calculation, except for approximations that are valid only in the deep type-II regime. Therefore, a simple way to locate the transition from type-I to type-II behavior seems to compute $H_c$ and $H_{c2}$ and determine 
the point at which $H_c=H_{c2}$. In a one-component system, this yields a critical value for the Ginzburg-Landau 
parameter $\kappa=\kappa_c=1/\sqrt{2}$, where $\kappa$ is the ratio
of magnetic penetration depth and coherence length. It turns out 
that at this point all three critical fields are identical, $H_c=H_{c1}=H_{c2}$. Then, for $\kappa>\kappa_c$ we have $H_{c2}>H_{c1}$ and flux tubes exist for magnetic fields between $H_{c2}$ and $H_{c1}$. This is the type-II regime. For $\kappa<\kappa_c$ there are no flux tubes and there is a first-order phase transition from the superconducting to the normal phase at the critical field $H_c$.
This is the type-I regime. An additional, but equivalent, criterion is 
the long-range interaction between flux tubes:
the interaction is repulsive for $\kappa>\kappa_c$ and attractive for $\kappa<\kappa_c$.

The situation is more complicated in a color superconductor. This was already realized in Refs.\ \cite{Giannakis:2003am,Iida:2004if}, where it was pointed out that various criteria for type-I/type-II behavior do not coincide, i.e., do not yield a single critical $\kappa$. 
A more detailed understanding of the transition region between type-I and type-II behavior was achieved in our recent general two-component study \cite{Haber:2017kth}, and we shall make use of the insights of this work. Moreover, in our present three-component system there is not simply a single superconducting phase and critical fields for the transition to the normal-conducting phase. 
Instead, we need to compute the critical fields for all possible transitions between the CFL, 2SC, and unpaired phases. 
Our strategy is thus as follows. We start with the homogeneous phases to construct a phase diagram at nonzero external 
magnetic field $H$. This corresponds to computing the various critical fields $H_c$. Then, we  compute the critical fields 
$H_{c2}$, and the intersection where $H_c=H_{c2}$ will give us an idea (although not a precise location, because of the multi-component structure) for the transition between type-I and type-II 
behavior. The resulting phase diagram is then used as a foundation for the calculation of the flux tube profiles and energies, which 
is done in the type-II regime. We will not attempt to resolve the details of the type-I/type-II transition region. This would require a fully numerical study of the flux tube lattice, as explained in more detail in Ref.\ \cite{Haber:2017kth}.

%%%%%%%%%%%%%%%%%%%%%%%%%%%%%%%%%%%%%%%%%%%%%%%%%%%%%%%%%%
\section{Homogeneous phases}
\label{sec:hom}
%%%%%%%%%%%%%%%%%%%%%%%%%%%%%%%%%%%%%%%%%%%%%%%%%%%%%%%%%%

We write the complex scalar fields as ($i=1,2,3$)
\be \label{rhopsi}
\phi_i = \frac{\rho_i}{\sqrt{2}}e^{i\psi_i} \, . 
\ee
In this section, we only consider homogeneous solutions, $\nabla\rho_i=\nabla\psi_i=0$ (then, the phases $\psi_i$ do not play any 
role). In this case, our ansatz for the gauge fields is ${\bf A}_3=xB_3{\bf e}_y$, $\tilde{\bf A}_8=x\tilde{B}_8{\bf e}_y$, such that the magnetic fields, given by the curl of the corresponding vector potentials, are homogeneous and parallel to the externally applied field ${\bf H}$. Then, the potential 
from Eq.\ (\ref{U123}) becomes
\bea \label{U0hom}
U_0 &=& \frac{B_3^2}{2}+\frac{\tilde{B}_8^2}{2}-\frac{\mu^2}{2}(\rho_1^2+\rho_2^2+\rho_3^2)+\frac{\lambda}{4}(\rho_1^4+\rho_2^4+\rho_3^4)-\frac{h}{2}(\rho_1^2\rho_2^2+\rho_1^2\rho_3^2+\rho_2^2\rho_3^2) \non[2ex]
&& +\frac{x^2\rho_1^2}{2}\left(\frac{g}{2}B_3+\tilde{g}_8\tilde{B}_8\right)^2+\frac{x^2\rho_2^2}{2}\left(-\frac{g}{2}B_3+\tilde{g}_8\tilde{B}_8\right)^2
+\frac{x^2\rho_3^2}{2} \left(2\tilde{g}_8\tilde{B}_8\right)^2 \, .
\eea
The equations of motion for ${\bf A}_3$ and $\tilde{\bf A}_8$ are
\begin{subequations} \label{A3A8}
\bea
0&=& \rho_1^2\left(\frac{g}{2}B_3+\tilde{g}_8\tilde{B}_8\right)-\rho_2^2\left(-\frac{g}{2}B_3+\tilde{g}_8\tilde{B}_8\right) \, , \\[2ex]
0&=& \rho_1^2\left(\frac{g}{2}B_3+\tilde{g}_8\tilde{B}_8\right)+\rho_2^2\left(-\frac{g}{2}B_3+\tilde{g}_8\tilde{B}_8\right)+4\rho_3^2\tilde{g}_8\tilde{B}_8  \, , 
\eea
\end{subequations}
and the equations of motion for the condensates $\rho_i$ are
\begin{subequations} \label{rho123}
\bea
0&=&\rho_1\left[\lambda\rho_1^2-h(\rho_2^2+\rho_3^2)-\mu^2+x^2\left(\frac{g}{2}B_3+\tilde{g}_8\tilde{B}_8\right)^2\right] \, , \label{rho1}\\[2ex]
0&=&\rho_2\left[\lambda\rho_2^2-h(\rho_1^2+\rho_3^2)-\mu^2+x^2\left(-\frac{g}{2}B_3+\tilde{g}_8\tilde{B}_8\right)^2\right] \, ,\label{rho2} \\[2ex]
0&=&\rho_3\left[\lambda\rho_3^2-h(\rho_1^2+\rho_2^2)-\mu^2+x^2\left(2\tilde{g}_8\tilde{B}_8\right)^2\right] \, .
\eea
\end{subequations}
Since in this section the condensates and magnetic fields are constant in space by assumption, the terms proportional to $x^2$ and the $x$-independent terms in Eqs.\ (\ref{rho123}) must vanish separately. As a consequence, the terms proportional to $x^2$ in the potential (\ref{U0hom}) vanish as well. This must be the case because  otherwise the free energy, obtained by integrating $U_0$ over space, would become infinite. We conclude that any given combination of nonzero condensates yields a condition for the magnetic fields. 
We discuss all possible combinations now.

\begin{itemize}

\item If all three condensates are nonzero, Eqs.\ (\ref{rho123}) show that $B_3=\tilde{B}_8=0$ [which trivially fulfills Eqs.\ (\ref{A3A8})]. This is the CFL solution, and Eqs.\  (\ref{rho123}) yield 
\be \label{rho0}
\rho_1^2=\rho_2^2=\rho_3^3 = \frac{\mu^2}{\lambda(1-2\eta)} \equiv \rho_{\rm CFL}^2 \, ,
\ee
where we have abbreviated the ratio of the cross-coupling constant to the self-coupling constant by
\be
\eta\equiv \frac{h}{\lambda} \, .
\ee
To ensure the boundedness of the potential, we must have $\eta<0.5$ (including all negative values), which also ensures 
$\rho_{\rm CFL}^2\ge 0$. 
With the weak-coupling results from Eq.\ (\ref{weak123}), $\eta=-0.5$. 
The Gibbs free energy density of the homogeneous CFL phase is now computed with the help of  Eqs.\ (\ref{GV}) and (\ref{U0hom}),
\be \label{GCFL}
\frac{G_{\rm CFL}}{V} = -\frac{H^2\cos^2\theta}{2} + U_{\rm CFL}  \, , 
\ee
where 
\be \label{UCFL}
U_{\rm CFL} = -\frac{3\mu^4}{4\lambda(1-2\eta)} \, .
\ee

\item If exactly one of the condensates vanishes, we also have $B_3=\tilde{B}_8=0$ in all three possible phases. The two non-vanishing condensates 
are identical, $\rho^2 = \mu^2/[\lambda (1-\eta)]$, and $U_0 = -\mu^4/[2\lambda(1-\eta)]$. We thus conclude that these phases are preferred over the CFL phase if and only if $\eta<-1$, for arbitrary magnetic field $H$. However, we shall see that in this regime the 2SC phase or the completely unpaired phase (to be discussed next) are preferred. Therefore, the phases in which exactly one of the three condensates is zero never occur and we  will ignore them from now on. 

\item If two of the condensates vanish, we have the following possible phases:

\begin{enumerate}

\item[$(i)$] $\rho_1=\rho_3=0$ ("2SC$_{\rm ud}$"). If we label the three color components as usual by (red, green, blue), this phase 
corresponds to Cooper pairing of red and blue up quarks with 
blue and red down quarks, respectively. In this case, Eqs.\ (\ref{A3A8})
yield a relation between $B_3$ and $\tilde{B}_8$, and Eq.\ (\ref{rho2}) yields the value for the nonzero condensate,
\be \label{rho2SC}
\rho_2^2 = \frac{\mu^2}{\lambda} \equiv \rho_{\rm 2SC}^2\, .
\ee 
Eliminating one of the magnetic fields, say $B_3$ in favor of $\tilde{B}_8$, in the Gibbs free energy (\ref{GV}) and minimizing the resulting expression with respect to $\tilde{B}_8$ yields
\be \label{2SC1}
\qquad B_3 = \frac{\sqrt{3}\sin\theta\cos\theta}{1+3\cos^2\theta} H \, , \qquad \tilde{B}_8 = \frac{3\sin\theta\cos^2\theta}{1+3\cos^2\theta} H \, , 
\ee
where we have used Eq.\ (\ref{g8}). The Gibbs free energy density becomes 
\be \label{G2SC}
\frac{G_{{\rm 2SC}_{\rm ud}}}{V} = -\frac{H^2\cos^2\theta}{2} -\frac{H^2}{2}\frac{3\sin^2\theta\cos^2\theta}{1+3\cos^2\theta}+ U_{{\rm 2SC}} \, , 
\ee
where
\be \label{U2SC}
U_{{\rm 2SC}} = -\frac{\mu^4}{4\lambda} \, .
\ee
Up to a relabeling of the colors [due to our flavor convention $Q={\rm diag}(1/3,1/3,-2/3)$], this phase is the phase  commonly termed 2SC in the literature. 
In the 2SC phase, we expect a Meissner effect for a certain combination of the photon 
and the eighth gluon, just like in CFL \cite{Schmitt:2003aa}. However, the result (\ref{G2SC}) shows that both $B_3$ and $\tilde{B}_8$ are nonzero. The reason is that 
the 2SC phase has a different mixing angle. Since we are interested in comparing the free energies of the different phases, we obviously have to work within the same basis for all phases. Our use of the CFL mixing angle, together with our convention for the charge matrix $Q$, 
therefore leads to a seemingly complicated result for the 2SC phase. The mixing angle of the 2SC phase can be recovered from these results by writing the Gibbs free energy (\ref{G2SC}) in the same form as the one for CFL (\ref{GCFL}),
\be \label{Gudhom}
\frac{G_{{\rm 2SC}_{\rm ud}}}{V} = -\frac{H^2\cos^2\vartheta_1}{2} + U_{{\rm 2SC}} \, , 
\ee
where
\be
\cos^2\vartheta_1 = \frac{3g^2}{3g^2+e^2} \, .
\ee
(In Sec.\ \ref{sec:2SC}, where we discuss defects in 2SC, we shall use an additional rotation given by $\vartheta_2$, hence the
notation $\vartheta_1$.)

\item[$(ii)$] $\rho_2=\rho_3=0$ ("2SC$_{\rm us}$"). This phase corresponds to green/blue and up/strange pairing. 
The only difference to the 2SC$_{\rm ud}$ phase is that $B_3$ has opposite sign, i.e., now ${\bf B}_3$
and $\tilde{\bf B}_8$ are anti-parallel, not parallel. In particular, the Gibbs free energies are identical, because $B_3$ enters quadratically. 
This is expected since we work in the massless limit and thus interchanging $d$ with $s$ quarks should not change any physics.

\item[$(iii)$]  $\rho_1=\rho_2=0$ ("2SC$_{\rm ds}$"). This phase  corresponds to red/green and down/strange pairing and is genuinely different from the usual 2SC phase -- even in the massless limit -- because now quarks with the 
same electric charge pair. In this case, we find $B_3=\tilde{B}_8=0$, $\rho_3^2=\mu^2/\lambda$, and
\be
\frac{G_{{\rm 2SC}_{\rm ds}}}{V} = -\frac{H^2\cos^2\theta}{2} + U_{{\rm 2SC}} \,  .
\ee
\end{enumerate}
Without magnetic field, these three phases have the same free energy and are preferred over the CFL phase for $\eta < -1$.
In the presence of a magnetic field, the Gibbs free energy of the 2SC$_{\rm ds}$ phase is always larger than that of the 2SC$_{\rm ud}$ and 2SC$_{\rm us}$ phases. Therefore, we no longer need to consider the 2SC$_{\rm ds}$ phase and use the term 2SC for both 2SC$_{\rm ud}$ and 2SC$_{\rm us}$ in the present section. (In Sec.\ \ref{sec:2SC} we will come back to the definitions of 2SC$_{\rm ud}$ and 2SC$_{\rm us}$ because we will discuss domain walls that interpolate between these two order parameters.)

\item Finally, in the completely unpaired phase ("NOR"), where $\rho_1=\rho_2=\rho_3=0$, we find 
\be
B_3=0 \, , \qquad \tilde{B}_8 = H\sin\theta \, , 
\ee
and the Gibbs free energy density is
\be
\frac{G_{\rm NOR}}{V}=-\frac{H^2}{2} \, .
\ee

\end{itemize}

\begin{figure} [t]
\begin{center}
\hbox{\includegraphics[width=0.5\textwidth]{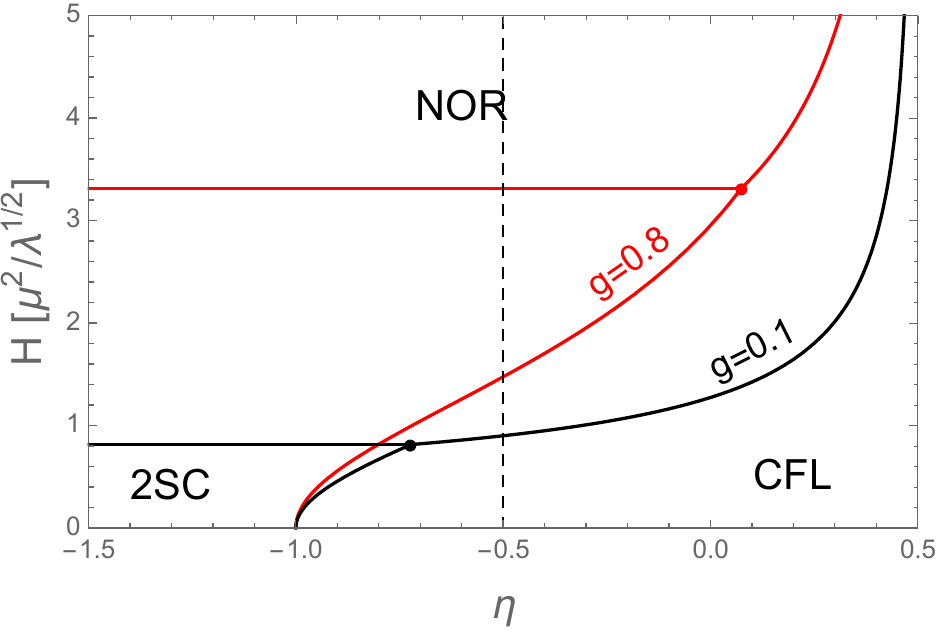}\includegraphics[width=0.5\textwidth]{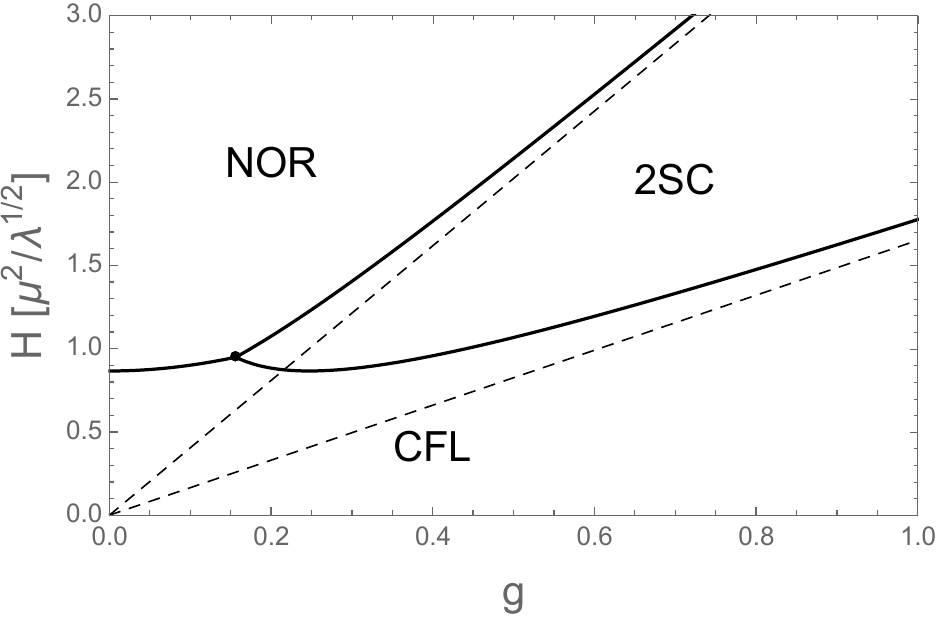}}
\caption{{\it Left panel:} phases in the plane of external magnetic field $H$ and ratio of cross-coupling to self-coupling $\eta=h/\lambda$. The solid lines are the critical fields $H_c$ from Eq.\ (\ref{Hcs}) for two different values of the strong coupling constant $g$. The vertical dashed line 
indicates the weak-coupling value $\eta=-0.5$. The horizontal scale terminates at the maximum value $\eta=0.5$, beyond which 
the Ginzburg-Landau potential becomes unbounded from below. {\it Right panel:} phases for $\eta=-0.5$. The critical point
where all three phases meet is given by $(g,H)=(2e/\sqrt{15},3/\sqrt{10}\,\mu^2/\sqrt{\lambda})$. For $g\to 0$, the critical 
field between CFL and NOR phases goes to $\sqrt{3}/2\, \mu^2/\sqrt{\lambda}$. The dashed lines are the critical fields for $g\gg e$.}
\label{fig:Heta}
\end{center}
\end{figure}

\subsection{Critical fields $H_{c}$}

With these results we can easily compute the critical magnetic fields of the phase transitions between CFL, 2SC, and NOR phases by comparing the corresponding free energies. We find
\be \label{Hcs}
\frac{H_c^2}{\mu^4/\lambda} = \left\{\begin{array}{cc} \displaystyle{\frac{3g^2+e^2}{2e^2}} & \mbox{2SC/NOR} \\[4ex] 
\displaystyle{\frac{3}{2(1-2\eta)}\frac{3g^2+4e^2}{4e^2}} & \mbox{CFL/NOR} \\[4ex]
\displaystyle{\frac{1+\eta}{1-2\eta}\frac{(3g^2+e^2)(3g^2+4e^2)}{9e^2g^2}} & \mbox{2SC/CFL} \end{array}\right. \, .
\ee
We plot the critical fields in the phase diagrams of Fig.\ \ref{fig:Heta}. In the chosen units for the magnetic field, the phase structure
only depends on $\eta$ and the strong coupling 
constant $g$ (the electromagnetic coupling constant $e$ is held fixed). This will no longer be
true when we discuss the type-I/type-II transition in the
subsequent sections. This transition  
depends also on $\lambda$ separately, i.e., on the ratio $T_c/\mu_q$. To avoid a multi-dimensional study of the parameter space, we shall thus later 
restrict ourselves to the weak-coupling results of the Ginzburg-Landau parameters, which imply $\eta=-0.5$, and extrapolate  these results 
to large values of $g$.  This is already done in the 
right panel of Fig.\ \ref{fig:Heta}, i.e., the left panel of this 
figure is the only plot where we keep $\eta$ general. 

We see that at zero magnetic field and weak coupling CFL is preferred over 2SC, which is well known and remains true if a small strange quark mass together with the conditions of color and electric neutrality are taken into account \cite{Alford:2002kj}. 
If $\eta$ is kept general, there is a regime where 2SC is preferred, even for vanishing magnetic field. This can be 
understood within the three-component picture, having in mind that $\eta=h/\lambda$ with $\lambda>0$: a negative coupling
$h$ implies repulsion between the three components. If this repulsion is sufficiently large, the condensates no longer "want" to coexist
and the 2SC phase becomes preferred.  

In the presence of a magnetic field $H$, the Gibbs free energy can be lowered by admitting this field into the system. 
In CFL, part of the magnetic field 
is already admitted because it is $\tilde{B}_8$, not $B$, that is completely expelled from the superconductor. Admitting 
a larger $B$ field can be achieved by breaking 
all condensates (now the entire applied magnetic field penetrates, $H=B$, but all condensation energy is lost) or by first going to the "intermediate" 2SC phase, where some condensation energy is maintained. Both scenarios are realized, as the right panel shows: 
for small values of the strong coupling constant, the CFL phase is directly superseded by the unpaired phase, while for all 
$g>2e/\sqrt{15}$ the 2SC phase appears between CFL and NOR.

\subsection{Critical fields $H_{c2}$}

Next, we compute the critical field $H_{c2}$ for all three phase transitions given in 
Eq.\ (\ref{Hcs}). We follow the standard procedure to compute these fields \cite{tinkham2004introduction}, which becomes slightly more complicated 
for the CFL/2SC transition, where we can follow the two-component treatment of Ref.\ \cite{Haber:2017kth}. 
The  equations of motion for the complex fields are computed from Eq.\ (\ref{U123}), 
\begin{subequations}
\bea
\left[\left(\nabla+i\frac{g}{2}{\bf A}_3+i\tilde{g}_8\tilde{\bf A}_8\right)^2+\mu^2-2\lambda|\phi_1|^2+2h(|\phi_2|^2+|\phi_3|^2)\right] \phi_1 &=& 0 \, , \\[2ex]
\left[\left(\nabla-i\frac{g}{2}{\bf A}_3+i\tilde{g}_8\tilde{\bf A}_8\right)^2+\mu^2-2\lambda|\phi_2|^2+2h(|\phi_1|^2+|\phi_3|^2)\right] \phi_2 &=& 0 \, , \\[2ex]
\left[\left(\nabla-2i\tilde{g}_8\tilde{\bf A}_8\right)^2+\mu^2-2\lambda|\phi_3|^2+2h(|\phi_1|^2+|\phi_2|^2)\right] \phi_3 &=& 0 \, .
\eea
\end{subequations}
We discuss the three phase transitions separately.

\begin{itemize}

\item The simplest case is the transition between 2SC and NOR, where $\phi_1=\phi_3=0$ in both phases. 
We linearize in $\phi_2$ and set ${\bf A}_3=0$ because $B_3=0$ in the unpaired phase.
This leaves the single equation 
\be
\left[\left(\nabla+i\tilde{g}_8\tilde{\bf A}_8\right)^2+\mu^2\right]\phi_2 = 0 \, .
\ee
With the usual argument \cite{tinkham2004introduction} this gives a maximal field $\tilde{B}_8=-\mu^2/\tilde{g}_8$.  
Since in the normal phase $\tilde{B}_8=H\sin\theta$, the critical field is
 \be
H_{c2} = \frac{3\mu^2}{e} \qquad \mbox{(for 2SC/NOR)} \, .
\ee
At the 2SC/NOR transition, the system is an ordinary single-component superconductor, and we expect an ordinary type-I/type-II transition at exactly $H_c=H_{c2}$. This can be confirmed by the numerical calculation of $H_{c1}$ for ordinary 2SC flux tubes, see Fig.\ \ref{fig:phases} in Sec.\ \ref{sec:results}. Therefore, using Eq.\ (\ref{Hcs}) and the 
weak-coupling expression for $\mu$ from Eq.\ (\ref{weak123}), 2SC flux tubes appear for
\be
\frac{T_c}{\mu_q}>\frac{\sqrt{7\zeta(3)}}{12\sqrt{3}\pi^2}\sqrt{g^2+\frac{e^2}{3}} \simeq 0.014 \sqrt{g^2+\frac{e^2}{3}} \, .
\ee
This standard type-I/type-II transition is expected to occur 
at $\kappa_{\rm 2SC}^2=1/2$. As a check, we may thus define the corresponding Ginzburg-Landau parameter a posteriori, 
\be
\kappa_{\rm 2SC}^2 = \frac{72\pi^4}{7\zeta(3)}\frac{3}{g^2+\frac{e^2}{3}} \frac{T_c^2}{\mu_q^2} \, , 
\ee
which is in exact agreement with Eq.\ (112) of Ref.\ \cite{Iida:2002ev}.

\item For the transition between CFL and NOR phases we linearize in all three condensates and set ${\bf A}_3=0$,
because in the phase above $H_{c2}$ all condensates and $B_3$ vanish. This leads to the three equations  
\bea
\left[\left(\nabla+i\tilde{g}_8\tilde{\bf A}_8\right)^2+\mu^2\right] \phi_1 =
\left[\left(\nabla+i\tilde{g}_8\tilde{\bf A}_8\right)^2+\mu^2\right] \phi_2 =
\left[\left(\nabla-2i\tilde{g}_8\tilde{\bf A}_8\right)^2+\mu^2\right] \phi_3 = 0 \, .
\eea
The first two equations give a maximal field $\tilde{B}_8=-\mu^2/\tilde{g}_8$, 
which we use to compute $H_{c2}$, such that at least one of the condensates is nonzero below $H_{c2}$. This definition of $H_{c2}$ for the CFL/NOR transition agrees with Ref.\ \cite{Iida:2004if}, and  we find the same critical field as for the 2SC/NOR transition,
 \be
H_{c2} = \frac{3\mu^2}{e}  \qquad \mbox{(for CFL/NOR)} \, .
\ee
As an estimate for the location of the type-I/type-II transition we again use the point $H_c=H_{c2}$, although in this case the critical 
region is expected to look more complicated because CFL is a multi-component system. We find that CFL flux tubes appear (if the next phase up in $H$ is the NOR phase) for
\be \label{Tmucond}
\frac{T_c}{\mu_q} > \frac{\sqrt{7\zeta(3)}}{24\pi^2}\frac{\sqrt{g^2+\frac{4}{3}e^2}}{\sqrt{1-2\eta}} \simeq 8.7\times 10^{-3}\sqrt{g^2+\frac{4}{3}e^2} \, ,
\ee
where, for the numerical estimate, we have set $\eta=-0.5$. 
As Fig.\ \ref{fig:Heta} demonstrates, the CFL/NOR transition is only relevant for $g<2e/\sqrt{15}\simeq  0.16$, where one would expect the weak-coupling results to be applicable. 
Hence, in this regime, $T_c/\mu_q \propto \exp(-{\rm const}/g)$ is exponentially suppressed and it seems very unlikely that the type-II regime is realized.

\item For the transition between CFL and 2SC, we use, without loss of generality, the 2SC$_{\rm ud}$ phase.  In this phase, $\phi_1=\phi_3=0$ and
thus we linearize in $\phi_1$ and $\phi_3$ (but not in $\phi_2$). Moreover, in 2SC$_{\rm ud}$ we have $g{\bf A}_3=2\tilde{g}_8\tilde{\bf A}_8$,
which follows from  Eq.\ (\ref{2SC1}). This relation is used to eliminate ${\bf A}_3$ and we arrive at the two equations
\be
\left[\left(\nabla\pm2i\tilde{g}_8\tilde{\bf A}_8\right)^2+\mu^2+2h|\phi_2|^2\right] \phi_{1/3} = 0 \, ,
\ee
and the homogeneous solution for the second condensate $|\phi_2|^2 = \mu^2/(2\lambda)$. With $\tilde{\bf A}_8 = x\tilde{B}_8{\bf e}_y$
this becomes 
\be
\mu^2(1+\eta)\phi_{1/3}=\left[-\Delta\mp 2i(2\tilde{g}_8\tilde{B}_8)x\partial_y+(2\tilde{g}_8\tilde{B}_8)^2x^2\right]\phi_{1/3} \, ,
\ee
where $\Delta = \partial_x^2+\partial_y^2+\partial_z^2$. As for the standard scenario, this equation has the form of the Schr{\"o}dinger 
equation for the harmonic oscillator, and we can compute the critical field in the usual way from the lowest eigenvalue
\cite{tinkham2004introduction,Haber:2017kth}. The result is
\be
H_{c2} = \frac{2\mu^2(1+\eta)(3g^2+e^2)}{3eg^2} \qquad \mbox{(for CFL/2SC)} \, .
\ee 
Again, we can determine the point $H_c=H_{c2}$, which suggests type-II behavior for
\be \label{Tcmu1}
\frac{T_c}{\mu_q} > \frac{\sqrt{14\zeta(3)}}{24\pi^2\sqrt{1-2\eta}\sqrt{1+\eta}}\frac{g\sqrt{3g^2+4e^2}}{\sqrt{3g^2+e^2}} \simeq 0.017 g\sqrt{\frac{3g^2+4e^2}{3g^2+e^2}} \,.
\ee
If we use the critical temperature for CFL from perturbative calculations \cite{Alford:2007xm,Schmitt:2002sc}, 
\be
T_c = 2^{1/3}\frac{e^\gamma}{\pi}\Delta_0 \, , 
\ee
with the Euler-Mascheroni constant $\gamma$ and the zero-temperature gap
\be
\Delta_0 = \mu_q b \exp\left(-\frac{3\pi^2}{\sqrt{2}g}\right) \, , \qquad 
b\equiv 512\pi^4\left(\frac{2}{g^2 N_f}\right)^{5/2}e^{-\frac{\pi^2+4}{8}}2^{-1/3} \, ,
\ee
and extrapolate the resulting ratio $T_c/\mu_q$ to large values of the coupling, we find that the criterion (\ref{Tcmu1}) for type-II behavior is not fulfilled for any $g$.
Thus, if we take Eq.\ (\ref{Tcmu1}) as the relevant criterion, we have to assume that strong-coupling effects, not captured by the extrapolation of the weak-coupling result, drive $T_c$ sufficiently large to allow for type-II behavior. 
As model calculations suggest, $T_c/\mu_q \gtrsim 0.06$ [choosing $g = 3.5$ in Eq.\ (\ref{Tcmu1}), which is plausible
for interiors of neutron stars] is not unrealistically large. We note, however, that the multi-component nature of CFL suggests that flux tubes can appear for {\it smaller} values of $T_c/\mu_q$ due to a possible first-order onset of flux tubes that
increases the region in the phase diagram where a lattice of flux tubes is preferred \cite{Haber:2017kth}. The exact calculation of the modified critical $T_c/\mu_q$ would require a numerical study of the flux tube lattice, and it is conceivable that even the extrapolated weak-coupling result allows for type-II behavior.

\end{itemize}

%%%%%%%%%%%%%%%%%%%%%%%%%%%%%%%%%%%%%%%%%%%%%%%%%%%%%%%%%%
\section{CFL flux tubes}
\label{sec:CFL}
%%%%%%%%%%%%%%%%%%%%%%%%%%%%%%%%%%%%%%%%%%%%%%%%%%%%%%%%%%

We now turn to the flux tube solutions in the CFL phase. The first step is the formulation of  the equations of motion in the most 
general way (within our diagonal ansatz for the gap matrix). This allows us to discuss the various possible flux tube 
configurations, compare their profiles and free energies, and determine the energetically 
most preferred flux tube configuration  by computing the critical fields $H_{c1}$. 

%%%%%%%%%%%%%%%%%%%%%%%%%%%%%%%%%%%%%%%%%%%%%%%%%%%%%%%%%%%%%%%%%%%%
\subsection{Equations of motion and flux tube energy}
%%%%%%%%%%%%%%%%%%%%%%%%%%%%%%%%%%%%%%%%%%%%%%%%%%%%%%%%%%%%%%%%%%%%

Having in mind a single, straight flux tube, we assume cylindrical symmetry and work in cylindrical coordinates ${\bf r} = (r,\varphi,z)$.  We write the modulus and the phase of the condensates from Eq.\ (\ref{rhopsi}) as  ($i=1,2,3$),
\be \label{fn}
\rho_i({\bf r}) = f_i(r)\rho_{\rm CFL} \, , \qquad \psi_i({\bf r}) = n_i\varphi \, , 
\ee
with the CFL condensate in the homogeneous phase $\rho_{\rm CFL}$ from Eq.\ (\ref{rho0}) and  
dimensionless functions $f_i(r)$. Single-valuedness of the order parameter requires $n_i\in \mathbb{Z}$. These are the winding numbers, for which there is a priori no additional condition, in particular they can be chosen independently of each other.  We will see that this choice determines the properties of the flux tube. 
For the gauge fields, we make the ansatz 
\be
{\bf A}_3({\bf r}) = \frac{a_3(r)}{r}{\bf e}_\varphi  \, , \qquad \tilde{\bf A}_8({\bf r}) = \frac{\tilde{a}_8(r)}{r}{\bf e}_\varphi  \, ,
\ee
with the dimensionless functions $a_3(r)$ and $\tilde{a}_8(r)$. This yields magnetic fields in the $z$ direction, 
\be \label{B38}
{\bf B}_3(r) = \frac{1}{r}\frac{\partial a_3}{\partial r}{\bf e}_z \, , \qquad \tilde{\bf B}_8(r) = \frac{1}{r}\frac{\partial \tilde{a}_8}{\partial r}{\bf e}_z
 \, .
 \ee
After eliminating $\mu$ in favor of $\rho_{\rm CFL}$ with the help of Eq.\ (\ref{rho0}), we can write the potential (\ref{U123}) as
\bea \label{U00}
U_0 &=& U_{\rm CFL} + U_{\circlearrowleft} \, , 
\eea
with $U_{\rm CFL}$ from Eq.\ (\ref{UCFL}) and the free energy density of the flux tube
\bea \label{Uflux}
U_{\circlearrowleft}&=&\frac{\lambda\rho_{\rm CFL}^4}{2}\left\{\frac{\lambda(a_3'^2+\tilde{a}_8'^2)}{R^2} + f_1'^2+f_2'^2+f_3'^2 +\frac{(1-f_1^2)^2}{2}+\frac{(1-f_2^2)^2}{2}+\frac{(1-f_3^2)^2}{2} \right.\non[2ex]
&& \left.+f_1^2\frac{{\cal N}_1^2}{R^2}+f_2^2\frac{{\cal N}_2^2}{R^2}+f_3^2\frac{{\cal N}_3^2}{R^2} -\eta\Big[(1-f_1^2)(1-f_2^2)+(1-f_1^2)(1-f_3^2)+(1-f_2^2)(1-f_3^2)\Big]\right\} \, , 
\eea
where we have introduced the new dimensionless coordinate 
\be
R=r\sqrt{\lambda}\,\rho_{\rm CFL} \, , 
\ee
have denoted derivatives with respect to $R$ by a prime, and have abbreviated
\be \label{N123}
{\cal N}_1 \equiv n_1+\frac{g}{2}a_3+\tilde{g}_8\tilde{a}_8 \, , \qquad {\cal N}_2 \equiv n_2-\frac{g}{2}a_3+\tilde{g}_8\tilde{a}_8
\, , \qquad {\cal N}_3\equiv n_3-2\tilde{g}_8\tilde{a}_8 \, .
\ee
Consequently, the equations of motion for the gauge fields become
\begin{subequations} \label{a3a8}
\bea
a_3''-\frac{a_3'}{R}&=&\frac{g}{2\lambda}\left(f_1^2{\cal N}_1-f_2^2{\cal N}_2\right) \, , \\[2ex]
\tilde{a}_8''-\frac{\tilde{a}_8'}{R}&=&\frac{\tilde{g}_8}{\lambda}\left(f_1^2{\cal N}_1+f_2^2{\cal N}_2
-2f_3^2{\cal N}_3\right) \, ,
\eea
\end{subequations}
and the equations of motion for the condensates are 
\begin{subequations} \label{f123}
\bea
0&=& f_1''+\frac{f_1'}{R}+f_1(1-f_1^2)-f_1\frac{{\cal N}_1^2}{R^2} -\eta f_1(2-f_2^2-f_3^2) \, , \\[2ex]
0&=& f_2''+\frac{f_2'}{R}+f_2(1-f_2^2)-f_2\frac{{\cal N}_2^2}{R^2} -\eta f_2(2-f_1^2-f_3^2) \, , \\[2ex]
0&=& f_3''+\frac{f_3'}{R}+f_3(1-f_3^2)-f_3\frac{{\cal N}_3^2}{R^2} -\eta f_3(2-f_1^2-f_2^2) \, . 
\eea
\end{subequations}
The boundary values of the scalar fields are as follows. Far away from the flux tube, the system is in the 
CFL phase, such that $f_i(\infty)=1$. In the origin, the scalar fields  vanish
if the respective component has nonzero winding, $f_i(0)=0$ if $n_i\neq 0$. Otherwise, we require $f_i'(0)=0$ as a boundary condition, and  $f_i(0)$ must be determined dynamically.
For the gauge fields, we use Eqs.\ (\ref{a3a8}) to determine their values at infinity. Assuming $a_3'(\infty)=a_3''(\infty)=\tilde{a}_8'(\infty)=\tilde{a}_8''(\infty)=0$, we find 
\be \label{ainf}
a_3(\infty) = \frac{n_2-n_1}{g} \, , \qquad \tilde{a}_8(\infty)=\frac{2n_3-n_1-n_2}{6\tilde{g}_8} \, .
\ee
In the origin we then have to require $a_3(0)=a_8(0)=0$, which follows from the equations of motion evaluated for small $R$. We solve the coupled differential equations (\ref{a3a8}) and (\ref{f123}) numerically 
with the help of a successive over-relaxation method to obtain the profiles of the flux tubes. The flux tube energy $F_\circlearrowleft$ 
per unit length is then 
obtained by inserting the result into Eq.\ (\ref{Uflux}) and integrating over space. We write the result as 
\bea \label{FL}
\frac{F_\circlearrowleft}{L} &=& \frac{1}{L}\int d^3{\bf r}\, U_\circlearrowleft 
= \pi\rho_{\rm CFL}^2 \,{\cal I}_\circlearrowleft \, , 
\eea
where $L$ is the length of the flux tube in the $z$-direction, and 
\bea \label{Icircle}
{\cal I}_\circlearrowleft\equiv \int_0^\infty dR\,R\left[\frac{\lambda(a_3'^2+\tilde{a}_8'^2)}{R^2}+\frac{1-f_1^4}{2}
+\frac{1-f_2^4}{2}+\frac{1-f_3^4}{2} -\eta(3-f_1^2f_2^2-f_1^2f_3^2-f_2^2f_3^2)\right] \, ,
\eea
where partial integration and the equations of motion (\ref{f123}) have been used. 

%%%%%%%%%%%%%%%%%%%%%%%%%%%%%%%%%%%%%%%%%%%%%%%%%%%%%%%%%%%%%%%%%%%%
\subsection{Critical field $H_{c1}$} 
\label{sec:Hc1}
%%%%%%%%%%%%%%%%%%%%%%%%%%%%%%%%%%%%%%%%%%%%%%%%%%%%%%%%%%%%%%%%%%%%

To determine the critical magnetic field $H_{c1}$ we need to compute the Gibbs free energy of the CFL phase 
in the presence of a flux tube. We insert the energy density $U_0$ from Eq.\ (\ref{U00}) with the notation introduced in 
Eq.\ (\ref{FL}) into the general form of the Gibbs free energy (\ref{GV}). Furthermore, 
we use 
\be \label{drB}
\int d^3{\bf r}\, \tilde{B}_8 = 2\pi L \tilde{a}_8(\infty) \, , 
\ee
which follows directly from the form of the magnetic field in Eq.\ (\ref{B38}) and the boundary condition $\tilde{a}_8(0)=0$. Recall that
we have defined $\tilde{\bf B}_8 = \tilde{B}_8 {\bf e}_z$, i.e., $\tilde{B}_8$ is the $z$-component, not the modulus, of $\tilde{\bf B}_8$. Therefore, ${\bf H}\cdot\tilde{\bf B}_8 =H \tilde{B}_8$ with $H$ being non-negative by assumption and the sign of $\tilde{B}_8$ 
indicating whether $\tilde{\bf B}_8$ is aligned or anti-aligned with ${\bf H}$. 

This yields the Gibbs free energy density 
\be
\frac{G}{V} = -\frac{H^2\cos^2\theta}{2} +U_{\rm CFL} +\frac{L}{V}\left[\frac{F_{\circlearrowleft}}{L}-2\pi \tilde{a}_8(\infty) H \sin\theta \right] \, .
\ee
It is favorable to place a single flux tube into the system if this reduces the free energy of the homogeneous CFL phase (\ref{GCFL}), i.e., if the expression in the square brackets becomes negative. By definition, 
this occurs at the critical magnetic field $H_{c1}$. Writing this critical field in the same units as the critical fields in Fig.\ \ref{fig:Heta}, we find 
\bea \label{Hc1}
\frac{H_{c1}}{\mu^2/\sqrt{\lambda}}  &=& \frac{(3g^2+4e^2)\,{\cal I}_\circlearrowleft}{4e\sqrt{\lambda}(1-2\eta)(n_1+n_2-2n_3)}
%=\frac{(3g^2+4e^2)}{4e\sqrt{\lambda}(1-2\eta)(n_1+n_2-2n_3)}\frac{F_\circlearrowleft}{\pi\rho^2_{CFL}L}
\, ,
\eea
where we have used Eqs.\ (\ref{costheta}), (\ref{g8}), (\ref{rho0}), (\ref{ainf}), and (\ref{FL}). Note that the critical field is proportional to the flux tube energy per winding number $n_1+n_2-2n_3$. In general, the expression on the right-hand side can be positive or negative, but we have assumed $H$ to be 
positive and hence $H_{c1}$ must be positive. We have $1-2\eta>0$ for all allowed values 
of $\eta$ and ${\cal I}_\circlearrowleft>0$ [which we always find to be the case, although it is not manifest from
Eq.\ (\ref{Icircle}) since $f_i(r)>1$ is possible]. Therefore,
the winding numbers must be chosen such that $n_1+n_2-2n_3>0$, which can be understood as follows. 
If $n_1+n_2-2n_3>0$, we have $\tilde{a}_8(\infty)<0$ because of Eq.\ (\ref{ainf}). 
Hence, due to $\tilde{a}_8(0)=0$ and Eq.\ (\ref{B38}), and assuming  $\tilde{a}_8(r)$ to be a monotonic function of $r$, $\tilde{\bf B}_8$ is {\it anti-parallel} to ${\bf H}$ for all $r$.  Therefore,  $\tilde{\bf B}_8\sin\theta$, which is the contribution to ${\bf B}$, is {\it parallel} to ${\bf H}$ because $\sin\theta<0$, as it should be. 

%%%%%%%%%%%%%%%%%%%%%%%%%%%%%%%%%%%%%%%%%%%%%%%%%%%%%%%%%%%%%%%%%%%%
\subsection{Asymptotic behavior}
\label{sec:asym}
%%%%%%%%%%%%%%%%%%%%%%%%%%%%%%%%%%%%%%%%%%%%%%%%%%%%%%%%%%%%%%%%%%%%

It is useful to determine the point at which the long-range interaction between two flux tubes 
changes from repulsive to attractive. In a multi-component system, this point is different from the point where $H_c=H_{c2}$ \cite{Haber:2017kth}. 
To compute the interaction between flux tubes, we first need to discuss the asymptotic behavior of the flux tube profiles. 
Far away from the center of the flux tube, i.e., for large $R$, we use the ansatz for the gauge fields $a_3(R) = a_3(\infty) + R v_3(R)$, $\tilde{a}_8(R) = \tilde{a}_8(\infty) + R \tilde{v}_8(R)$ and for the scalar fields 
$f_i(R)=1+u_i(R)$ ($i=1,2,3$). We assume $n_1+n_2+n_3=0$. This is equivalent to a vanishing baryon circulation far away from the 
flux tube, as will be discussed in detail in Sec.\ \ref{sec:circ}.  

We linearize the equations of motion (\ref{a3a8}) and (\ref{f123}) in the functions $v_3, \tilde{v}_8, u_1, u_2, u_3$. 
The equations for the gauge fields then yield decoupled equations for $v_3$ and $\tilde{v}_8$, 
\begin{subequations}
\bea
v_3''+\frac{v_3'}{R} &\simeq & \left(1+\frac{R^2}{\kappa_3^2}\right) \frac{v_3}{R^2} \, , \\[2ex]
\tilde{v}_8''+\frac{\tilde{v}_8'}{R} &\simeq & \left(1+\frac{R^2}{\tilde{\kappa}_8^2}\right) \frac{\tilde{v}_8}{R^2} \, , 
\eea
\end{subequations}
where we have used Eq.\ (\ref{ainf}), and where
\bea \label{kappa38}
\kappa_3^2 \equiv \frac{2\lambda}{g^2} \, , \qquad \tilde{\kappa}_8^2\equiv\frac{\lambda}{6\tilde{g}_8^2} \, .
\eea
The solutions of these equations are 
\begin{subequations} \label{asymp}
\bea
v_3(R) &=& c_3K_1(R/\kappa_3) \, , \\[2ex]
\tilde{v}_8(R) &=& \tilde{c}_8K_1(R/\tilde{\kappa}_8) \, , 
\eea
\end{subequations}
where $K_n$ are the modified Bessel functions of the second kind and $c_3$ and $\tilde{c}_8$ are integration constants which can 
only be determined numerically.
The linearized equations for the scalar fields are 
\begin{subequations} 
\bea
0&\simeq& u_1''+\frac{u_1'}{R}-2u_1+2\eta(u_2+u_3) \, , \\[2ex]
0&\simeq& u_2''+\frac{u_2'}{R}-2u_2+2\eta(u_1+u_3) \, , \\[2ex]
0&\simeq& u_3''+\frac{u_3'}{R}-2u_3+2\eta(u_1+u_2) \, .
\eea
\end{subequations}
We solve these coupled equations by first writing them as
\bea
\Delta u = M u \, , \qquad M\equiv 2\left(\begin{array}{ccc}1&-\eta&-\eta\\-\eta&1&-\eta\\-\eta&-\eta&1\end{array}\right) \, , \qquad u \equiv \left(\begin{array}{c} u_1\\u_2\\u_3\end{array}\right) \, ,
\eea
where $\Delta$ is the Laplacian in cylindrical coordinates.
This system of equations can be diagonalized, 
\be
\Delta\tilde{u} = (U^{-1}MU) \tilde{u} \, , 
\ee
with $\tilde{u}=U^{-1}u$ and 
\be
U = \left(\begin{array}{ccc} 1&-1&-1\\1&0&1\\1&1&0 \end{array}\right) \, , \qquad U^{-1}MU = \left(\begin{array}{ccc}\nu_1&0&0\\0&\nu_2&0\\0&0&\nu_2\end{array}\right) \, ,
\ee
where the eigenvalues of $M$ are denoted by 
\be
\nu_1\equiv 2(1-2\eta) \, , \qquad \nu_2\equiv 2(1+\eta) \, .
\ee
Solving the uncoupled equations and then undoing the rotation yields the 
asymptotic solutions 
\begin{subequations} \label{asympf}
\bea
u_1(R)&=&d_1K_0(\sqrt{\nu_1}R)-(d_2+d_3)K_0(\sqrt{\nu_2}R) \, , \\[2ex]
u_2(R)&=&d_1K_0(\sqrt{\nu_1}R)+d_3K_0(\sqrt{\nu_2}R) \, , \\[2ex]
u_3(R)&=&d_1K_0(\sqrt{\nu_1}R)+d_2K_0(\sqrt{\nu_2}R) \, , 
\eea
\end{subequations}
with  integration constants $d_1$, $d_2$, $d_3$. From Fig.\ \ref{fig:Heta} we know that the CFL phase only exists for $-1<\eta<0.5$. For values outside that regime the 2SC phase is preferred (large negative values of $\eta$), or the Ginzburg-Landau potential is unbounded from below 
(large positive values). Therefore, both eigenvalues $\nu_1$ and $\nu_2$ are positive in the relevant regime and the square roots
in Eqs.\ (\ref{asympf}) are real. 

We have thus found that all gauge fields and scalar fields fall off exponentially for 
$R\to\infty$, which guarantees the finiteness of the free energy of the flux tube configuration and justifies the boundary conditions used above for the gauge fields. This is not the case if the baryon circulation is nonzero, $n_1+n_2+n_3\neq 0$, where, as suggested from 
ordinary superfluid vortices, at least one of the fields falls off with a power law \cite{Eto:2009kg}.

%%%%%%%%%%%%%%%%%%%%%%%%%%%%%%%%%%%%%%%%%%%%%%%%%%%%%%%%%%%%%%%%%%%%
\subsection{Interaction between flux tubes}
\label{sec:inter}
%%%%%%%%%%%%%%%%%%%%%%%%%%%%%%%%%%%%%%%%%%%%%%%%%%%%%%%%%%%%%%%%%%%%

We can now use the asymptotic solutions to compute the interaction between two flux tubes at large distances. 
This calculation has been explained in detail for a two-component system in Ref.\ \cite{Haber:2017kth}, based on 
well-known approximations for a one-component superconductor \cite{Kramer:1971zza}.
 The extension to the present case with three scalar components 
and two gauge fields is straightforward, although somewhat tedious. The interaction energy $F^\circlearrowleft_{\rm int}(R_0)$
between two flux tubes, say flux tube $(a)$ and flux tube $(b)$, whose centers are in a distance $R_0$ from each other, is defined as 
\be
F^{(a)+(b)}= F^{(a)}+ F^{(b)}+ F^\circlearrowleft_{\rm int}(R_0) \, ,
\ee
where $F^{(a)+(b)}$ is the total free energy of the two flux tubes, $F^{(a)}$ is the free energy of flux tube $(a)$ in the absence of flux 
tube $(b)$, and vice versa for $F^{(b)}$.  We give a brief sketch of the calculation in appendix \ref{app:inter}. The result 
for the interaction energy per unit length is 
\bea \label{Fint}
\frac{F_{\rm int}^{\circlearrowleft}}{L} &=& 2\pi \rho_{\rm CFL}^2\left[\frac{\kappa_3^2g^2c_3^2}{2}K_0(R_0/\kappa_3)
+6\tilde{\kappa}_8^2\tilde{g}_8^2\tilde{c}_8^2K_0(R_0/\tilde{\kappa}_8)-3d_1^2K_0(\sqrt{\nu_1}R_0)-2(d_2^2+d_3^2+d_2d_3)K_0(\sqrt{\nu_2}R_0)\right] \, . \;\;
\eea
This is in agreement with Eq.\ (46) in Ref.\ \cite{Iida:2004if}, where the term proportional to $K_0(R_0/\kappa_3)$ was absent because 
only flux tubes without $B_3$-flux were considered. 
There are positive (repulsive) contributions from the gauge fields and negative (attractive) contributions from the scalar 
fields. For $\eta < 0$ we have $\nu_2<\nu_1$, and thus the long-distance behavior of the attractive contribution is dominated by $K_0(\sqrt{\nu_2}R_0)$ [note that $2(d_2^2+d_3^2+d_2d_3)=(d_2+d_3)^2+d_2^2+d_3^2>0$]. Since at weak coupling $\eta=-0.5$, we shall focus on this case. 
For the repulsive part we notice that always $\kappa_3 >\tilde{\kappa}_8$, such that, if there is a nonzero $B_3$-flux, the dominant contribution is 
given by $K_0(R_0/\kappa_3)$. Then, the interaction is attractive for $\sqrt{\nu_2}<1/\kappa_3$. If the $B_3$-flux vanishes, the contribution containing $\kappa_3$ does not exist and the 
interaction is attractive for  $\sqrt{\nu_2}<1/\tilde{\kappa}_8$. Inserting the definitions for $\kappa_3$ and $\tilde{\kappa}_8$ from Eq.\ (\ref{kappa38}),
we find that the interaction is repulsive for 
\bea \label{repul}
\frac{T_c}{\mu_q} > \left\{\begin{array}{cc} \displaystyle{\frac{\sqrt{7\zeta(3)}}{12\pi^2\sqrt{2(1+\eta)}}\sqrt{g^2+\frac{4}{3}e^2}\simeq 0.025\sqrt{g^2+\frac{4}{3}e^2}} & \;\;\;\;\mbox{for $B_3=0$} \\ [4ex]
\displaystyle{\frac{\sqrt{7\zeta(3)}}{12\pi^2\sqrt{2(1+\eta)}}g\simeq 0.025 g } & \;\;\;\;\mbox{for $B_3\neq0$} \end{array}\right. \, , 
\eea
where, for the numerical approximation, we have inserted the weak-coupling result $\eta=-0.5$. We shall make use of these results in our discussion of the phase diagram in Sec.\ \ref{sec:results}.

%%%%%%%%%%%%%%%%%%%%%%%%%%%%%%%%%%%%%%%%%%%%%%%%%%%%%%%%%%%%%%%%%%%%
\subsection{Baryon circulation and magnetic flux}
\label{sec:circ}
%%%%%%%%%%%%%%%%%%%%%%%%%%%%%%%%%%%%%%%%%%%%%%%%%%%%%%%%%%%%%%%%%%%%

In general, the flux tubes described by Eqs.\ (\ref{a3a8}) and (\ref{f123})  have nonzero baryon circulation $\Gamma$ and 
nonzero magnetic fluxes $\Phi_3$ and $\tilde{\Phi}_8$. We use these three quantities to discuss the properties of the possible flux tube 
configurations.

The baryon circulation is computed by inserting our ansatz for the order parameter into the superfluid velocity (\ref{superv}) to obtain
\bea \label{VS}
{\bf v}_s 
&=& \frac{1}{6\mu_q}\frac{\rho_1^2 n_1+\rho_2^2n_2+\rho_3^2n_3+\tilde{g}_8\tilde{a}_8(\rho_1^2+\rho_2^2-2\rho_3^2)+\frac{g}{2}a_3(\rho_1^2-\rho_2^2)}{\rho_1^2+\rho_2^2+\rho_3^2} \frac{{\bf e}_\theta}{r}
\, ,
\eea
where we have used $u^2=1/3$. Then, the baryon circulation around a CFL flux tube along a circle at infinity becomes
\be
\Gamma = \oint d{\bm\ell}\cdot{\bf v}_s = \frac{\pi}{3\mu_q}\frac{n_1+n_2+n_3}{3} \, ,
\ee
where we have used that far away from the flux tube the condensates assume their homogeneous CFL values and become identical, 
$\rho_1=\rho_2=\rho_3$. 
Consequently, the CFL flux tube has vanishing baryon circulation if the three winding numbers add up to zero. In particular, 
the gauge fields have dropped out of the result. This is different from an ordinary flux tube in a single-component superconductor, where
the circulation can only vanish due to a cancellation between the winding number and the gauge field, as can be seen by setting $\rho_1=\rho_2=0$ in Eq.\ (\ref{VS}). 

The magnetic fluxes are
\begin{subequations}\label{Phi3Phi8}
\bea
\Phi_3 &=& \oint d{\bm \ell} \cdot {\bf A}_3 = 2\pi a_3(\infty) = \frac{2\pi}{g}(n_2-n_1)  \, , \\[2ex]
\tilde{\Phi}_8 &=& \oint d{\bm \ell} \cdot \tilde{\bf A}_8 = 2\pi \tilde{a}_8(\infty) = \frac{\pi}{\tilde{g}_8}\frac{2n_3-n_1-n_2}{3}  \, . 
\label{Phi8}
\eea
\end{subequations} 
We can now classify all possible flux tubes by their three winding numbers and use the baryon circulation and the color-magnetic 
fluxes to understand their main properties. In Table \ref{tab:n1n2n3} we list the most important configurations that are expected to 
appear in CFL in the presence of an externally imposed  rotation and/or an externally imposed  magnetic field. One point 
of this table is to demonstrate that the CFL line defects considered so far in the literature and the new configurations discussed 
here are all defined by a particular choice of the triple of winding numbers.  (We recall that the three-component nature of our 
system is a consequence of the diagonal ansatz of the gap matrix. In principle, more components might appear through non-diagonal gap matrices, which would induce additional color magnetic fields.  To our knowledge, such configurations have not been studied in the literature.) 

\begin{table}[t]
\begin{center}
\begin{tabular}{|c|c||c|c|c|} 
\hline
\rule[-1.5ex]{0em}{4ex} 
 CFL line defect  &$\;\;$ $(n_1,n_2,n_3)$ $\;\;$ & $\;\;$ $\Gamma\; [\pi/3\mu_q]$ $\;\;$& $\;\;$$\Phi_3\; [2\pi/g]$ $\;\;$
  & $\;\;$$\tilde{\Phi}_8\; [\pi/\tilde{g}_8]$ $\;\;$ \\[1ex] \hline\hline
\rule[-1.5ex]{0em}{6ex} 
$\;\;$ $T_{111}$ (global vortex \cite{Forbes:2001gj})$\;\;$& $(n,n,n)$ & $n$ & 0 & 0\\[2ex] \hline
\rule[-1.5ex]{0em}{6ex} 
$T_{001}$ (semi-superfluid vortex, "$M_1$" \cite{Balachandran:2005ev}) & $(0,0,n)$ & $\displaystyle{\frac{n}{3}}$ & 0 & $\displaystyle{\frac{2n}{3}}$ \\[2ex] \hline
\rule[-1.5ex]{0em}{6ex} 
$T_{110}$ (semi-superfluid vortex, "$M_2$" \cite{Balachandran:2005ev}) & $(n,n,0)$ & $\displaystyle{\frac{2n}{3}}$ & 0 & $\displaystyle{-\frac{2n}{3}}$ \\[2ex] \hline
\rule[-1.5ex]{0em}{6ex} 
$T_{112}$ (magnetic flux tube \cite{Iida:2004if}) & $(n,n,-2n)$ & 0 & 0 & $-2n$  \\[2ex] \hline
\rule[-1.5ex]{0em}{6ex} 
$\;\;$$T_{101}$ (magnetic flux tube, new in this work)$\;\;$ & $(n,0,-n)$ & 0 & $-n$ & $-n$  \\[2ex] \hline
\end{tabular}
\caption{Line defects in CFL, classified by the  winding numbers of the three components of the order parameter, 
$n\in \mathbb{Z}$,  from which baryon number circulation $\Gamma$ and color-magnetic fluxes $\Phi_3$ and $\tilde{\Phi}_8$ are obtained. 
}
\label{tab:n1n2n3}
\end{center}
\end{table}

If an external rotation is applied to CFL, vortices with nonzero 
baryon circulation must be formed. This has been discussed in detail in the literature. For instance, it has been found that the global vortex $T_{111}$ (which has no color-magnetic flux) is unstable with respect to decay into three so-called semi-superfluid vortices
\cite{Balachandran:2005ev,Alford:2016dco}. 
Each semi-superfluid vortex has nonzero color-magnetic fluxes, but a triple of vortices $T_{100}$, $T_{010}$, $T_{001}$
(in an obvious generalization of the notation introduced in Table \ref{tab:n1n2n3}) is color neutral. We do not discuss rotationally induced 
vortices here. We rather focus on configurations with vanishing baryon circulation $\Gamma$ and non-vanishing 
magnetic flux $\tilde{\Phi}_8$,
\begin{subequations} \label{constraints}
\bea
n_1+n_2+n_3&=&0 \, , \\[2ex]
n_1+n_2-2n_3& >& 0 \, .
\eea
\end{subequations}
These are flux tubes that are formed in the type-II regime of CFL if an external (ordinary) 
magnetic field is applied, but no rotation. 
In the interior of a neutron star, there is nonzero rotation {\it and} a nonzero magnetic field, i.e., the total magnetic flux and the 
total angular momentum must be nonzero. We know that the rotational axis and the magnetic field axis are, at least for some neutron stars,  not aligned, otherwise we would not observe them as pulsars. This suggests that, if there is a CFL core in the pulsar,  magnetic flux and baryon circulation are 
not maintained by a single species of flux tubes. Therefore, it appears that purely magnetic flux tubes, without circulation, are necessary.  

\begin{figure} [t]
\begin{center}
\hbox{\includegraphics[width=0.5\textwidth]{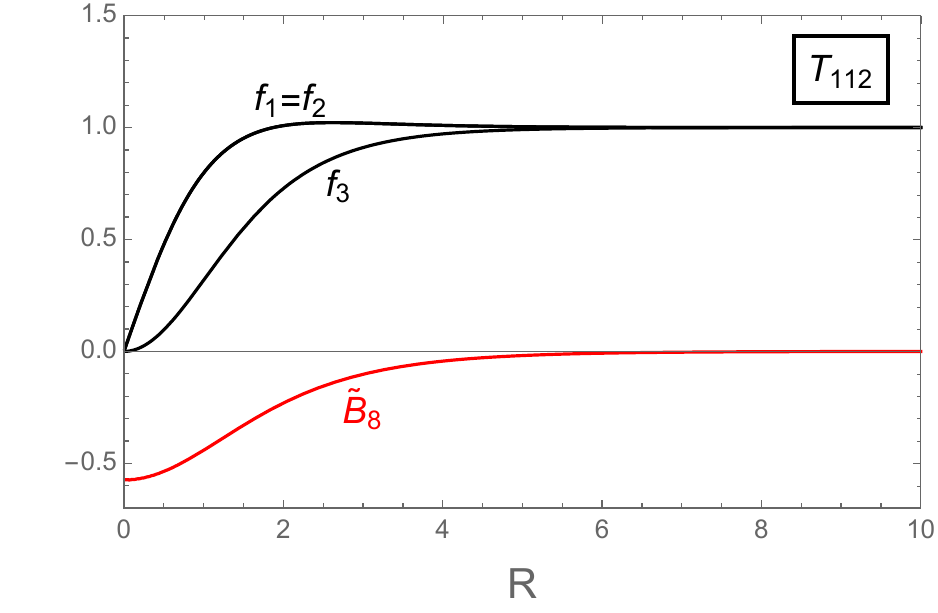}\includegraphics[width=0.5\textwidth]{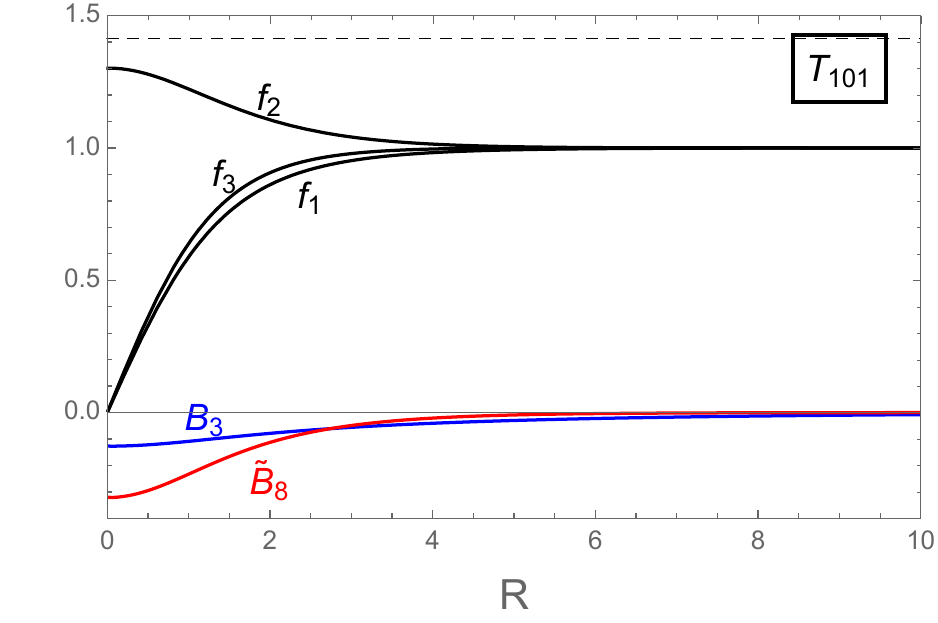}}
\caption{Dimensionless condensates $f_1$, $f_2$, $f_3$ and magnetic fields $B_3$ and $\tilde{B}_8$ in units 
of $\mu^2/\sqrt{\lambda}$ for the CFL flux tubes $T_{112}$ (left, where $B_3=0$) and $T_{101}$ (right) with $n=1$ at the same coupling $g=0.1$ and $\eta=-0.5$, 
$T_c/\mu_q\simeq 0.012$, as a function of the dimensionless radial coordinate $R=r\sqrt{\lambda}\,\rho_{\rm CFL}$. The horizontal dashed line in the right panel marks the homogeneous 2SC condensate $f_2=\sqrt{1-2\eta}$ ($f_2$
is rescaled with the CFL condensate). If we increase the winding, $n_1, -n_3\to \infty$, the condensate $f_2$  approaches this value. The negative sign of $\tilde{B}_8$ ensures that $\tilde{B}_8\sin\theta \,{\bf e}_z$ is aligned with the 
magnetic field ${\bf H}$. At the relatively small value of $g$ chosen here, the $B_3$ field in the right panel falls off on a larger length 
scale than the $B_8$ field, $\kappa_3/\tilde{\kappa}_8=\sqrt{1+4e^2/(3g^2)}\simeq 3.6$. }
\label{fig:profilesCFL}
\end{center}
\end{figure}

Within the two constraints (\ref{constraints})  we are interested in the energetically most preferred 
flux tube. In the previous literature, only the flux tube $T_{112}$ was discussed, but there are obviously 
infinitely many more possibilities to choose  winding numbers that fulfill the constraints (\ref{constraints}). 
One can systematically study all possibilities: for instance, define the length (squared) of the vector $(n_1,n_2,n_3)$ by 
$N^2\equiv n_1^2+n_2^2+n_3^2$, then choose an $N_0$ and solve the equations of motion for all vectors $(n_1,n_2,n_3)$ 
that fulfill Eqs.\ (\ref{constraints}) and whose length is smaller than $N_0$. This can easily be automatized with a computer. 
We have done such a calculation and have compared the free energies of the different flux tubes (for a certain choice of the 
Ginzburg-Landau parameters). The result suggests that the obvious expectation is fulfilled: unless we are in the type-I regime, where flux tubes are never preferred, configurations with 
a small "total winding" $N$ tend to be favored. Therefore, we do not go into the details of this analysis, and focus exclusively 
on the two configurations with the smallest $N$, namely $T_{112}$ and $T_{101}$.  

The price one has to pay for minimizing 
the total winding in $T_{101}$ compared to $T_{112}$ is a nonzero $B_3$ field. This gives an energy cost due to the $B_3^2$ term in the 
free energy. (Presumably this is the reason why this flux tube has so far been ignored in the literature.) However, one of the 
scalar fields has zero winding and thus it is allowed to remain nonzero in the center of the flux tube. 
Moreover, the negative sign of the effective coupling constant $h$ (using the weak-coupling result) suggests that the scalar components interact repulsively with each other. Hence, if 
$\rho_1$ and $\rho_3$ go to zero, $\rho_2$ does not only not vanish, but is even expected to be enhanced in the center of the flux tube. This implies a gain in condensation energy and is exactly what our numerical result will show. 

There is another way of understanding the difference between $T_{101}$ and $T_{112}$. 
If, in the configuration $T_{112}$, the winding $n$ is increased, the flux tube gets wider and the completely unpaired phase 
in the center of the tube grows until eventually CFL has been replaced by the NOR phase. As a consequence, $H_{c1}$ approaches $H_c$ for $n\to \infty$. (In the type-I regime, $H_{c1}\to H_c$ from above, and in the type-II regime from below.) This suggests that, in the absence of flux tubes, there is a transition from the CFL to the NOR phase. However, we have seen in 
Sec.\ \ref{sec:hom} that there is a parameter regime where CFL is, upon increasing $H$, replaced by 2SC, not by the NOR phase. 
The configuration $T_{101}$ accounts for this transition: now, if the winding $n$ is sent to infinity, the second component survives and one 
arrives in the 2SC phase (more precisely, the 2SC$_{\rm ud}$ phase). This suggests that where there is a transition from CFL to 2SC, 
the configuration $T_{101}$ should be favored. 

We will thus refer to $T_{112}$ as a "CFL flux tube with a NOR core" and to $T_{101}$
as a "CFL flux tube with a 2SC core", keeping in mind that this is a simplifying terminology for the fully dynamically computed flux tube profiles. 
We show the profiles of both configurations in Fig.\ \ref{fig:profilesCFL} for the coupling constant $g=0.1$ and the ratio $T_c/\mu_q$ at which the critical fields $H_{c1}$ of both configurations turn out to be identical. 
We shall compare the critical magnetic field $H_{c1}$ for both kind of flux tubes more systematically in Sec.\ \ref{sec:results}.

%%%%%%%%%%%%%%%%%%%%%%%%%%%%%%%%%%%%%%%%%%%%%%%%%%%%%%%%%%
\subsection{Physical units and numerical estimates}
\label{sec:units}
%%%%%%%%%%%%%%%%%%%%%%%%%%%%%%%%%%%%%%%%%%%%%%%%%%%%%%%%%%

As already pointed out in Refs.\ \cite{Iida:2001pg,Iida:2004if}, the critical magnetic fields associated with the (partial) breaking of color superconductivity
are extremely large. The main reason is that color superconductors -- in an astrophysical environment where $g\gg e$ -- admit 
a large part of the externally applied magnetic field because the massless gauge boson is almost identical to the photon, 
with a small admixture of one of the gluons. Therefore, breaking the superconductor, or partially breaking it through the formation 
of  magnetic defects, requires an enormously large ordinary magnetic field. In all our results, the magnetic fields are given 
in units of $\mu^2/\sqrt{\lambda}$, which is very convenient since it minimizes the number of parameters we have to specify. To translate this into physical units we use the definitions (\ref{weak123}) and find
\be \label{HinG}
\frac{\mu^2}{\sqrt{\lambda}}  \simeq 1.597\times 10^{19}(1-t)\mu_{q400}^2\frac{T_c}{\mu_q}\, G \, ,
\ee 
where $t\equiv T/T_c$ and $\mu_{q400}\equiv \mu_q/(400\, {\rm MeV})$. Although the ratio $T_c/\mu_q$ is exponentially small at weak coupling, this is certainly not true in the interior of neutron stars. 
Therefore, Eq.\ (\ref{HinG}) shows that the critical magnetic fields (for instance in Fig.\ \ref{fig:Heta})  are 
much larger than the measured magnetic fields at the surface of the star, which are at most of the order of
$10^{16} \,  {\rm G}$. Magnetic fields in the interior that are several orders of magnitude larger seem unlikely, although not inconceivable,
 given the estimate of maximal magnetic fields in a quark matter core of the order of $10^{20} \, {\rm G}$ \cite{Ferrer:2010wz}. 
As we shall see later, the new flux tube solution $T_{101}$ has a smaller critical field $H_{c1}$ compared to $T_{112}$,
but this decrease does  not change the order of magnitude estimate of the critical field strength.

We may also estimate the width of the flux tubes in physical units. From the 
asymptotic solutions of the CFL flux tubes (\ref{asympf}) and the definition of the dimensionless radial coordinate 
$R=r \sqrt{\lambda}\,\rho_{\rm CFL}$ we read off the coherence length $\xi$. This is the length scale on which all three 
condensates approach their homogeneous values. Again using Eqs.\ (\ref{weak123}) we find 
\be
\xi^{-1} = \sqrt{\lambda}\,\rho_{\rm CFL}  \simeq 10.76 \frac{T_c}{\mu_q}\sqrt{1-t}\,\mu_{q400} \, {\rm fm}^{-1} \, .
\ee
For a numerical estimate, let us set $T_c\simeq 40\, {\rm MeV}$, such that $T_c/\mu_q \simeq 0.1$. Judging from 
model calculations and extrapolations from the perturbative result, this is a large, but conceivable, critical temperature. 
Then, setting $T=0$, we find that $\xi \simeq 0.93 \, {\rm fm}$. 
The penetration depth $\ell$, i.e., the scale on which the magnetic fields fall off, is obtained from the asymptotic solution 
(\ref{asymp}). We have to distinguish between the penetration depths of $B_3$ and $\tilde{B}_8$, which become identical only for 
$g\gg e$,
\begin{subequations}
\bea
\ell_3^{-1} &=& \frac{g\rho_{\rm CFL}}{\sqrt{2}} \simeq 0.37 g\sqrt{1-t}\,\mu_{q400} \, {\rm fm}^{-1} \, , \\[2ex]
\ell_8^{-1} &=&\sqrt{6}\tilde{g}_8\rho_{\rm CFL} \simeq 0.37 \sqrt{g^2+\frac{4e^2}{3}} \sqrt{1-t}\,\mu_{q400} \, {\rm fm}^{-1}  \, .
\eea
\end{subequations}
With $T=0$ and $g\simeq 3.5$ we find $\ell_3\simeq \ell_8 \simeq 0.77\, {\rm fm}$. 

Finally, we write the energy of the flux tube per unit length from Eq.\ (\ref{FL}) as 
\be
\frac{F_{\circlearrowleft}}{L} = \pi\rho_{\rm CFL}^2 {\cal I}_{\circlearrowleft} \simeq 1.378\times 10^{9}(1-t)\mu_{q400}^2\,{\cal I}_{\circlearrowleft}\,\frac{\rm erg}{\rm cm} \, , 
\ee
where ${\cal I}_{\circlearrowleft}$ has to be computed numerically. For instance, with $g=3.5$ and $T_c/\mu_q=0.1$ we find
for the $T_{112}$ tube ${\cal I}_{\circlearrowleft} \simeq 5.9$ and for the $T_{101}$ tube ${\cal I}_{\circlearrowleft} \simeq 2.5$, both 
with $n=1$, in rough agreement with the simple estimates used in Ref.\ \cite{Glampedakis:2012qp}, which yield $F_{\circlearrowleft}/L\simeq 1.5 \times 10^{10}\mu_{q400}^2\,{\rm erg}/{\rm cm}$.

%%%%%%%%%%%%%%%%%%%%%%%%%%%%%%%%%%%%%%%%%%%%%%%%%%%%%%%%%%
\section{2SC flux tubes and domain walls}
\label{sec:2SC}
%%%%%%%%%%%%%%%%%%%%%%%%%%%%%%%%%%%%%%%%%%%%%%%%%%%%%%%%%%

At first sight, color-magnetic flux tubes in 2SC (= flux tubes that approach the 2SC phase at infinity) are 
less exotic than their counterparts in CFL because 2SC is a single-component
superconductor, i.e., only one of the scalar fields in the Ginzburg-Landau potential is nonzero. In an ordinary 2SC flux tube, which 
we will refer to as\footnote{To distinguish 2SC flux tubes from CFL flux tubes, we denote them by $S$, instead of $T$. The 2SC domain wall will be denoted by $D$.} $S_1$, this component has a nonzero winding and vanishes in the center of the tube \cite{Alford:2010qf}. 
One may ask, however, whether the other two components are induced inside the flux tube, similarly to the flux tubes discussed in 
Refs.\ \cite{Forgacs:2016ndn,Forgacs:2016iva}. We shall investigate this 
possibility by considering 2SC flux tubes within the full three-component calculation. The result suggests the 
existence of domain walls, which will emerge as the infinite-radius limit of the flux tubes.

In the 2SC phase, we work with $Q={\rm diag}(2/3,-1/3,-1/3)$, i.e., we order the quark flavors as $(u,d,s)$. Then, the usual 2SC phase with 
up/down pairing, 2SC$_{\rm ud}$,  is given by a nonzero condensate $\rho_3$. Since we work in the massless limit, this phase is
equivalent to the 2SC$_{\rm us}$ phase, where only $\rho_2$ is nonzero\footnote{Recall that in all preceding sections we used $Q={\rm diag}(-1/3,-1/3,2/3)$, which is more convenient for CFL, and thus the 2SC$_{\rm us}$ and 2SC$_{\rm ud}$ phases were given by 
a nonzero $\rho_1$ and $\rho_2$, respectively.}. For the magnetic defects in 2SC, it is convenient to introduce the following rotated fields\footnote{In many aspects, the 2SC calculation is analogous to the CFL calculation, and it is helpful to reflect this in the notation. We have therefore decided to write 
$\tilde{A}_\mu^8$ and 
$\tilde{A}_\mu$ again, although these fields are different from the rotated fields in the CFL calculation. Since the CFL mixing will not appear from now on, this should not lead to any confusion.},
\be
\left(\begin{array}{c} \tilde{A}_\mu^3 \\ \tilde{A}_\mu^8 \\ \tilde{A}_\mu \end{array}\right) =  
\left(\begin{array}{ccc} \cos\vartheta_2 & 0 &\sin\vartheta_2 \\ 0&1&0 \\ -\sin\vartheta_2&0&\cos\vartheta_2  \end{array}\right)\left(\begin{array}{ccc} 1 & 0 &0 \\ 0& \cos\vartheta_1 & \sin\vartheta_1 \\ 0 & -\sin\vartheta_1 & \cos\vartheta_1   \end{array}\right)\left(\begin{array}{c} A_\mu^3 \\ A_\mu^8 \\ A_\mu \end{array}\right)  \, ,
\ee
with 
\begin{subequations}
\bea
\sin\vartheta_1 &=& \frac{e}{\sqrt{3g^2+e^2}} \, , \qquad \cos\vartheta_1 = \frac{\sqrt{3}g}{\sqrt{3g^2+e^2}} \, , \\[2ex]
\sin\vartheta_2 &=& \frac{\sqrt{3}e}{\sqrt{3g^2+4e^2}} \, , \qquad \cos\vartheta_2 = \frac{\sqrt{3g^2+e^2}}{\sqrt{3g^2+4e^2}} \, .
\eea
\end{subequations}
This two-fold rotation is motivated as follows. If we were interested in the homogeneous 2SC$_{\rm ud}$ 
phase, given by a nonzero $\rho_3$, the gauge field $A_\mu^3$ 
would play no role and applying the rotation given by $\vartheta_1$  yields a magnetic field that is expelled, $\tilde{B}_8$, and the orthogonal combination that penetrates the 2SC phase. This is well-known, see for instance Ref.\ \cite{Schmitt:2003aa}. 
Here, however,  we are interested in keeping all condensates. One finds that $\rho_1$ and $\rho_2$ are charged under all 
three gauge fields that are obtained from this first rotation. The second rotation,  given by $\vartheta_2$, simplifies the situation by
creating a field, namely $\tilde{A}_\mu$, under which {\it all three} condensates are neutral, while leaving $\tilde{A}_\mu^8$ unchanged.  
This is useful because it eliminates $\tilde{A}_\mu$ from the calculation of the flux tube and domain wall profiles, and we only have to 
deal with two gauge fields in the numerical calculation. 

The Ginzburg-Landau potential in terms of the new rotated fields is obtained by starting from the potential given by Eqs.\ (\ref{UUB}) and (\ref{U123}), undoing the CFL rotation and applying the 2SC rotations, or by re-starting from the original potential (\ref{UPhi}). In either case, one derives
\be
U = \frac{\tilde{B}^2}{2} + U_0 \, , 
\ee
with 
\bea \label{U02SC}
U_0 &=& \frac{\tilde{\bf B}_3^2}{2}+ \frac{\tilde{\bf B}_8^2}{2} +\frac{(\nabla\rho_1)^2}{2}+\frac{(\nabla\rho_2)^2}{2}+\frac{(\nabla\rho_3)^2}{2}-\frac{\mu^2}{2}(\rho_1^2+\rho_2^2+\rho_3^2)+\frac{\lambda}{4}(\rho_1^4+\rho_2^4+\rho_3^4) -\frac{h}{2}(\rho_1^2\rho_2^2+\rho_2^2\rho_3^2+\rho_1^2\rho_3^2)\non[2ex]
&&+\left(\nabla\psi_1+\tilde{q}_3\tilde{\bf A}_3+\tilde{q}_{81}\tilde{\bf A}_8\right)^2\frac{\rho_1^2}{2}
+\left(\nabla\psi_2-\tilde{q}_3\tilde{\bf A}_3+\tilde{q}_{82}\tilde{\bf A}_8\right)^2\frac{\rho_2^2}{2}+\left(\nabla\psi_3-\tilde{q}_{83}\tilde{\bf A}_8\right)^2\frac{\rho_3^2}{2} \, ,
\eea
where we have written the scalar fields in terms of their moduli and phases according to Eq.\ (\ref{rhopsi}), and where 
we have abbreviated 
\be
\tilde{q}_{81}\equiv \frac{3g^2+4e^2}{6\sqrt{3g^2+e^2}} \, , \qquad \tilde{q}_{82}\equiv \frac{3g^2-2e^2}{6\sqrt{3g^2+e^2}} \, , \qquad \tilde{q}_{83}\equiv \frac{\sqrt{3g^2+e^2}}{3} \, , 
\ee 
and
\be
\tilde{q}_3\equiv \frac{g}{2}\sqrt{\frac{3g^2+4e^2}{3g^2+e^2}} \, .
\ee
We can write the Gibbs free energy density as
\be
\frac{G}{V} = -\frac{H^2\cos^2\vartheta_1\cos^2\vartheta_2}{2}+\frac{1}{V}\int d^3{\bf r}
\left[U_0-H(\tilde{B}_3\cos\vartheta_1\sin\vartheta_2+\tilde{B}_8\sin\vartheta_1)\right] \, ,
\ee
where we have used $\tilde{B} = H\cos\vartheta_1\cos\vartheta_2$, which follows from minimizing $G$ with respect to $\tilde{B}$. 
For the homogeneous phases we repeat the calculation from Sec.\ \ref{sec:hom} to find
\begin{subequations} \label{B3B823}
\bea
&{\rm 2SC}_{\rm ud}:& \qquad \tilde{B}_3=H\cos\vartheta_1\sin\vartheta_2 \, , \qquad \tilde{B}_8 = 0 \, ,\\[2ex]
&{\rm 2SC}_{\rm us}:& \qquad \tilde{B}_3=\frac{3ge(3g^2-2e^2)H}{2\sqrt{3g^2+4e^2}(3g^2+e^2)^{3/2}} \, , \qquad \tilde{B}_8 = \frac{9g^2eH}{2(3g^2+e^2)^{3/2}} \, . 
\eea
\end{subequations}

%%%%%%%%%%%%%%%%%%%%%%%%%%%%%%%%%%%%%%%%%%%%%%%%%%%%%%%%%%
\subsection{Flux tubes in 2SC}
%%%%%%%%%%%%%%%%%%%%%%%%%%%%%%%%%%%%%%%%%%%%%%%%%%%%%%%%%%

In analogy to the CFL calculation, we write the scalar fields as 
$\rho_i({\bf r}) = f_i(r)\rho_{\rm 2SC}$ with the homogeneous 2SC condensate $\rho_{\rm 2SC}$ from Eq.\ (\ref{rho2SC}), and introduce the winding numbers in the 
phases through $\psi_i({\bf r}) = n_i\varphi$. We use the 2SC$_{\rm ud}$ phase 
for our boundary condition far away from the flux tube, i.e., 
$f_1(\infty)=f_2(\infty)=0$, $f_3(\infty)=1$, while $f_i(0)=0$ if the corresponding winding number $n_i$ is nonzero.
For the gauge fields we write 
\be \label{A3A8tt}
 \tilde{\bf A}_3({\bf r}) = \left[\frac{H\cos\vartheta_1\sin\vartheta_2}{2}r+\frac{\tilde{a}_3(r)}{r}\right]{\bf e}_\varphi \, ,
 \qquad \tilde{\bf A}_8({\bf r}) = \frac{\tilde{a}_8(r)}{r}{\bf e}_\varphi \, ,
\ee
with $\tilde{a}_3'(\infty)=\tilde{a}_8'(\infty)=0$ and $\tilde{a}_3(0)=\tilde{a}_8(0)=0$. In contrast to the CFL flux tubes, there is a magnetic field, $\tilde{B}_3$, which is nonzero far away from the 
flux tube (in addition to the homogeneous field $\tilde{B}$, which simply penetrates the superconductor). This field 
will become inhomogeneous in the flux tube, unless the system chooses to keep $\rho_1$ and $\rho_2$ zero
everywhere. We have separated the 
homogeneous part of the $\tilde{B}_3$ field in our ansatz (\ref{A3A8tt}), such that far away from the flux tube $\tilde{a}_3$ does not contribute to the magnetic field and we have $\tilde{\bf B}_3(\infty)=H\cos\vartheta_1\sin\vartheta_2\, {\bf e}_z$. 
This separation is useful, but not crucial. Alternatively,  one could have implemented the external field in the boundary condition for $\tilde{a}_3$.

Inserting our ansatz into the potential (\ref{U02SC}), we compute the Gibbs free energy density
\bea
&&U-{\bf H}\cdot {\bf B} = U_{\rm 2SC} -\frac{H^2\cos^2\vartheta_1}{2} - \lambda\rho_{\rm 2SC}^2H\sin\vartheta_1\frac{\tilde{a}_8'}{R} 
+ \frac{\lambda\rho_{\rm 2SC}^4}{2}\left\{\frac{\lambda(\tilde{a}_3'^2+\tilde{a}_8'^2)}{R^2}+f_1'^2+f_2'^2+f_3'^2+f_1^2\left(\frac{f_1^2}{2}-1\right) \right.\non[2ex]
&&\left. +f_2^2\left(\frac{f_2^2}{2}-1\right)+\frac{(1-f_3^2)^2}{2} +\frac{({\cal N}_1+\Xi R^2)^2f_1^2+({\cal N}_2-\Xi R^2)^2f_2^2+{\cal N}_3^2f_3^2}{R^2} -\eta(f_1^2f_2^2+f_1^2f_3^2+f_2^2f_3^2)\right\} \, , 
\eea
with $U_{\rm 2SC}$ from Eq.\ (\ref{U2SC}). Analogously to Sec.\ \ref{sec:CFL} we have introduced the dimensionless coordinate 
$R=r\sqrt{\lambda}\,\rho_{\rm 2SC}$, prime denotes derivative with respect to $R$, we have defined the dimensionless external magnetic field 
\be \label{Xi}
\Xi=\frac{\tilde{q}_3 H\cos\vartheta_1\sin\vartheta_2}{2\lambda\rho_{\rm 2SC}^2} = \frac{3eg^2}{4\sqrt{\lambda}(3g^2+e^2)}\frac{H}{\mu^2/\sqrt{\lambda}} \, ,
\ee
and we have abbreviated 
\bea
{\cal N}_1 &\equiv& n_1+\tilde{q}_3\tilde{a}_3+\tilde{q}_{81}\tilde{a}_8 \, ,\qquad 
{\cal N}_2 \equiv n_2-\tilde{q}_3\tilde{a}_3+\tilde{q}_{82}\tilde{a}_8 \, ,\qquad 
{\cal N}_3 \equiv n_3-\tilde{q}_{83}\tilde{a}_8 \, , 
\eea
in analogy to Eq.\ (\ref{N123}). 

The equations of motion for the gauge fields are
\begin{subequations}
\bea
\tilde{a}_3''-\frac{\tilde{a}_3'}{R}&=& \frac{\tilde{q}_3}{\lambda}[({\cal N}_1+\Xi R^2)f_1^2-({\cal N}_2-\Xi R^2)f_2^2] \, , \\[2ex]
\tilde{a}_8''-\frac{\tilde{a}_8'}{R}&=& \frac{1}{\lambda}[\tilde{q}_{81}({\cal N}_1+\Xi R^2)f_1^2+\tilde{q}_{82}({\cal N}_2-\Xi R^2)f_2^2-\tilde{q}_{83}{\cal N}_3f_3^2] \, , \label{a8t}
\eea
\end{subequations}
and for the scalar fields we find
\begin{subequations}
\bea
0&=& f_1''+\frac{f_1'}{R}+f_1\left[1-f_1^2-\frac{({\cal N}_1+\Xi R^2)^2}{R^2} +\eta(f_2^2+f_3^2)\right] \, , \\[2ex]
0&=& f_2''+\frac{f_2'}{R}+f_2\left[1-f_2^2-\frac{({\cal N}_2-\Xi R^2)^2}{R^2} +\eta(f_1^2+f_3^2)\right] \, , \\[2ex]
0&=& f_3''+\frac{f_3'}{R}+f_3\left[1-f_3^2-\frac{{\cal N}_3^2}{R^2} +\eta(f_1^2+f_2^2)\right] \, . 
\eea
\end{subequations}
Evaluating Eq.\ (\ref{a8t}) at $R=\infty$ yields 
\be \label{n3a}
\tilde{a}_8(\infty) = \frac{n_3}{\tilde{q}_{83}} \, ,
\ee
which is the usual relation for a single-component superconductor and implies vanishing baryon circulation far away from the flux tube.
There is no analogous condition for $\tilde{a}_3(\infty)$, and we determine this value dynamically in the numerical solution. 

We can write the Gibbs free energy density as 
\be \label{G2SCflux}
\frac{G}{V}=U_{\rm 2SC}-\frac{H^2\cos^2\vartheta_1}{2}+\frac{L}{V}\left[\frac{F_\circlearrowleft}{L}-2\pi \tilde{a}_8(\infty)H\sin\vartheta_1\right] \, , 
\ee
where the flux tube energy per unit length, in analogy to the CFL calculation, is
\be
\frac{F_\circlearrowleft}{L} = \pi\rho_{\rm 2SC}^2\,  {\cal I}_\circlearrowleft \, , 
\ee
with
\be
{\cal I}_\circlearrowleft \equiv \int_0^\infty dR\,R\left[\frac{\lambda(\tilde{a}_3'^2+\tilde{a}_8'^2)}{R^2}-\frac{f_1^4}{2}-\frac{f_2^4}{2}+\frac{1-f_3^4}{2}+\eta
(f_1^2f_2^2+f_1^2f_3^2+f_2^2f_3^2)\right] \, .
\ee
The critical magnetic field $H_{c1}$ is again calculated by setting the expression in the square brackets in Eq.\ (\ref{G2SCflux}) to zero, since the remaining terms are the Gibbs free energy density of the homogeneous 2SC phase (\ref{Gudhom}). However, this calculation is more complicated than in the CFL phase because $F_\circlearrowleft$ now depends implicitly on $H$. Therefore, instead of simply computing the free energy of the flux tube we have to solve  the following equation numerically,
\be \label{Xictube}
\Xi_{c1} - \frac{g^2 {\cal I}_\circlearrowleft(\Xi_{c1})}{8\lambda n_3} = 0 \, .
\ee
 In the simple case of the ordinary 2SC flux tube, i.e., where only the condensate $\rho_3$ is nonzero and where only the gauge field $\tilde{a}_8$ needs 
to be taken into account in the calculation of profiles, the free energy of the flux tube does not depend on the external magnetic field. In this case, it is useful to write Eq.\ (\ref{Xictube}) in the form
\be
\frac{H_{c1}}{\mu^2/\sqrt{\lambda}} = \frac{(3g^2+e^2){\cal I}_\circlearrowleft}{6e\sqrt{\lambda}n_3}= \frac{(3g^2+e^2)}{6e\sqrt{\lambda}n_3}\int_0^\infty dR\, R\left(\frac{1-f_3^4}{2}+\l\frac{\tilde{a}_8'^2}{R^2}\right)  \, , 
\ee 
where now the right-hand side directly yields the critical magnetic field.

%%%%%%%%%%%%%%%%%%%%%%%%%%%%%%%%%%%%%%%%%%%%%%%%%%%%%%%%%%
\subsection{Domain walls in 2SC}
\label{sec:domain}
%%%%%%%%%%%%%%%%%%%%%%%%%%%%%%%%%%%%%%%%%%%%%%%%%%%%%%%%%%

The profiles of the flux tubes from the previous subsection approach the 2SC$_{\rm ud}$ phase at infinity. We know that in the massless limit considered here the 2SC$_{\rm us}$ phase is equivalent to the 2SC$_{\rm ud}$ phase. Therefore, we can construct a 
domain wall that
approaches 2SC$_{\rm us}$ far away from the wall on one side and 2SC$_{\rm ud}$ on the other side. It is conceivable that the "twist" 
that changes 2SC$_{\rm us}$ into 2SC$_{\rm ud}$ admits a magnetic field in the wall, which leads to a gain in Gibbs free energy 
and might favor the domain wall over the homogeneous phase in the presence of an externally applied field. We shall see that 
this is indeed the case and that, in a certain parameter regime,  the domain wall solution is 
favored over the flux tubes from the previous subsection. 

Domain walls in the 2SC phase in the presence of a magnetic field were already suggested in Ref.\ \cite{Son:2007ny}. These domain 
walls are associated with the axial $U(1)_A$. This symmetry is broken due to the axial anomaly 
of QCD, but becomes an approximate symmetry at high density and is spontaneously broken by the 2SC condensate. These domain walls are perpendicular to the magnetic field 
and their width is given by the inverse of the mass of the $U(1)_A$ pseudo-Goldstone boson. This is different from the domain walls discussed here, which align themselves parallel to the magnetic field and which have finite width even though our potential does not 
include $U(1)_A$ breaking terms. The "anomalous" 
domain walls have been discussed within an effective Lagrangian for the Goldstone mode \cite{Son:2007ny}, and it would be interesting
for future work to investigate their competition or coexistence with the domain walls discussed here in a common framework.

The equations that have to be solved to compute the profile of 
the domain wall are derived as follows. Due to the geometry of the problem, we  work in cartesian coordinates 
rather than the cylindrical coordinates
used for the flux tubes. We keep the external magnetic field in the $z$-direction and, without loss of generality,  place the domain wall in the $y$-$z$-plane, such that the problem becomes one-dimensional along the $x$-axis. 
For the gauge fields, our ansatz is
\bea
\tilde{\bf A}_3({\bf r}) = \left[(x-x_0)H\cos\vartheta_1\sin\vartheta_2+ \sqrt{\lambda}\rho_{\rm 2SC}\tilde{a}_3(x)\right]{\bf e}_y \, , \qquad 
\tilde{\bf A}_8({\bf r})= \sqrt{\lambda}\rho_{\rm 2SC} \tilde{a}_8(x){\bf e}_y \, , 
\eea
such that the magnetic fields point in the $z$-direction with $z$-components
\be \label{BBB}
\tilde{B}_3 = H\cos\vartheta_1\sin\vartheta_2 +\lambda\rho_{\rm 2SC}^2\tilde{a}_3' \, , \qquad 
\tilde{B}_8=\lambda\rho_{\rm 2SC}^2\tilde{a}_8' \, , 
\ee
where prime now denotes the derivative with respect to the dimensionless coordinate $X\equiv \sqrt{\lambda}\rho_{\rm 2SC}\, x$. 
We have added an $x$-independent term proportional to $x_0$ to the gauge field $\tilde{\bf A}_3$. This term is irrelevant for the 
magnetic field and does not affect any physics. It is merely a useful term for the numerical evaluation because it can be used to shift
 the location of the domain wall on the $x$-axis. Since this location depends on the values of the parameters, we conveniently 
 adjust $x_0$ to keep the domain wall in the $x$-interval which we have chosen for the numerical calculation.

We set $\rho_1=0$ and introduce the dimensionless condensates as above through $\rho_i({\bf r})=f_i(x)\rho_{\rm 2SC}$ for $i=2,3$. 
As just explained, the phases of the condensates do not wind as we move across the wall, and thus we set $\psi_i=0$. 
One could define a new angle $\alpha$ by writing $f_1= f \cos\alpha$, $f_2= f \sin\alpha$ and solve the equations of motion for $f$ and $\alpha$, see Ref.\ \cite{Chernodub:2010sg} for a similar calculation in a two-component superconductor. This angle, which 
rotates between the two condensates, {\it does} wind across the domain wall. 
But this change of basis is not necessary, and we
 shall stick to the variables $f_1$, $f_2$.  Then, from Eq.\ (\ref{U02SC}) we compute the Gibbs free energy density 
\bea
U-{\bf H}\cdot {\bf B} &=& U_{\rm 2SC} -\frac{H^2\cos^2\vartheta_1}{2} - \lambda\rho_{\rm 2SC}^2H\sin\vartheta_1\,\tilde{a}_8' 
+ \frac{\lambda\rho_{\rm 2SC}^4}{2}\Big\{\lambda(\tilde{a}_3'^2+\tilde{a}_8'^2)+f_2'^2+f_3'^2 \non[2ex]
&&+\left[{\cal M}_2-2\Xi(X-X_0)\right]^2f_2^2+{\cal M}_3^2f_3^2-f_2^2-f_3^2+\frac{1}{2}(f_2^4+f_3^4)+\frac{1}{2}-\eta f_2^2f_3^2 \Big\} \, , 
\eea
with $\Xi$ from Eq.\ (\ref{Xi}), $X_0\equiv \sqrt{\lambda}\rho_{\rm 2SC} \,x_0$, and
\be
{\cal M}_2\equiv -\tilde{q}_3\tilde{a}_3+\tilde{q}_{82}\tilde{a}_8 \, , \qquad 
{\cal M}_3\equiv -\tilde{q}_{83}\tilde{a}_8 \, . 
\ee
The equations of motion are
\begin{subequations}
\bea
\tilde{a}_3'' &=& -\frac{\tilde{q}_3}{\lambda}\left[{\cal M}_2-2\Xi(X-X_0)\right]f_2^2 \, , \label{aX}\\[2ex]
\tilde{a}_8''&=& \frac{\tilde{q}_{82}}{\lambda} \left[{\cal M}_2-2\Xi(X-X_0)\right]f_2^2 -\frac{\tilde{q}_{83}}{\lambda}{\cal M}_3 f_3^2 \, ,
\eea
\end{subequations} 
and
\begin{subequations}
\bea
0&=& f_2'' + f_2\left\{1-f_2^2-\left[{\cal M}_2-2\Xi(X-X_0)\right]^2+\eta f_3^2\right\} \, , \label{f2X}\\[2ex]
0&=& f_3'' + f_3\left(1-f_3^2-{\cal M}_3^2+\eta f_2^2\right) \, . \label{f3X}
\eea
\end{subequations}
The boundary conditions are determined as follows. On one side far away from the domain wall, say at $X=+\infty$,  we put the 
2SC$_{\rm ud}$ phase, while on the other side, at $X=-\infty$, we put the 2SC$_{\rm us}$ phase. Then, the boundary conditions for the scalar fields are $f_2(+\infty)=f_3(-\infty)=0$ and $f_2(-\infty)=f_3(+\infty)=1$.
For the boundary conditions of the gauge fields we need the magnetic fields of the two phases far away from the wall 
(\ref{B3B823}) to find
\bea \label{bound2SC2}
\tilde{a}_3'(-\infty) &=& -\frac{4\tilde{q}_3\Xi}{g^2} \, , \qquad \tilde{a}_8'(-\infty) = \frac{6\Xi}{\sqrt{3g^2+e^2}} \, ,  \qquad 
\tilde{a}_3'(+\infty)=\tilde{a}_8'(+\infty)=0 \, .
\eea
Here the external field $H$ appears inevitably in the boundary conditions (in its dimensionless version $\Xi$),
while this was avoided in the case of the flux tubes by separating the $H$-dependent part in the ansatz for $\tilde{\bf A}_3$.
In addition to the boundary conditions for the derivatives, we have $\tilde{a}_8(+\infty)=0$, which follows from evaluating 
Eq.\ (\ref{f3X}) at $X=+\infty$. All other boundary values of the gauge fields must be determined dynamically.

The Gibbs free energy density becomes
\bea \label{GVdomain}
\frac{G}{V}&=&U_{\rm 2SC}-\frac{H^2\cos^2\vartheta_1}{2}+\frac{A_{yz}}{V}\frac{\sqrt{\lambda}\rho_{\rm 2SC}^3}{2}{\cal I}_{||} \, , 
\eea
where $A_{yz}$ is the area of the system in the plane of the domain wall, and the dimensionless energy per unit area of the domain wall is, after partial integration and using the equations of motion, 
\be
{\cal I}_{||}  \equiv \int_{-\infty}^\infty dX \left[\lambda(\tilde{a}_3'^2+\tilde{a}_8'^2)
-4\lambda\Xi\frac{\tan\vartheta_1}{\tilde{q}_3\sin\vartheta_2}\tilde{a}_8' +\frac{1}{2}(1-f_2^4-f_3^4)+\eta f_2^2f_3^2\right] \, .
\ee
As a check, we confirm that the integrand goes to zero at $X=\pm\infty$: the contribution of the scalar fields is obviously zero 
at $X=\pm\infty$ because one of the two functions $f_2$ and $f_3$ goes to 0 and the other one to 1. The gauge field contribution at $X=+\infty$ is obviously zero  because all derivatives  $\tilde{a}_3'$, $\tilde{a}_8'$ vanish. At $X=-\infty$, we employ the 
boundary conditions from Eq.\ (\ref{bound2SC2}) to show that the contributions quadratic in the derivatives of the gauge field 
are exactly canceled by the term proportional to $\tilde{a}_8'$. This term comes from the
${\bf H}\cdot {\bf B}$ term in the Gibbs free energy and was written separately in the flux tube energies in the previous sections, see for instance Eq.\ (\ref{G2SCflux}). Since here, in the case of the domain walls,  this would have required writing down a 
divergent integral [with the divergence being canceled by the divergent $\tilde{a}_8(-\infty)$],  
we have included the term linear in $\tilde{a}_8'$ into the integral.

%%%%%%%%%%%%%%%%%%%%%%%%%%%%%%%%%%%%%%%%%%%%%%%%%%%%%%%%%%
\subsection{Numerical results and discussion of profiles}
\label{sec:profiles}
%%%%%%%%%%%%%%%%%%%%%%%%%%%%%%%%%%%%%%%%%%%%%%%%%%%%%%%%%%

\begin{figure} [t]
\begin{center}
\hbox{\includegraphics[width=0.5\textwidth]{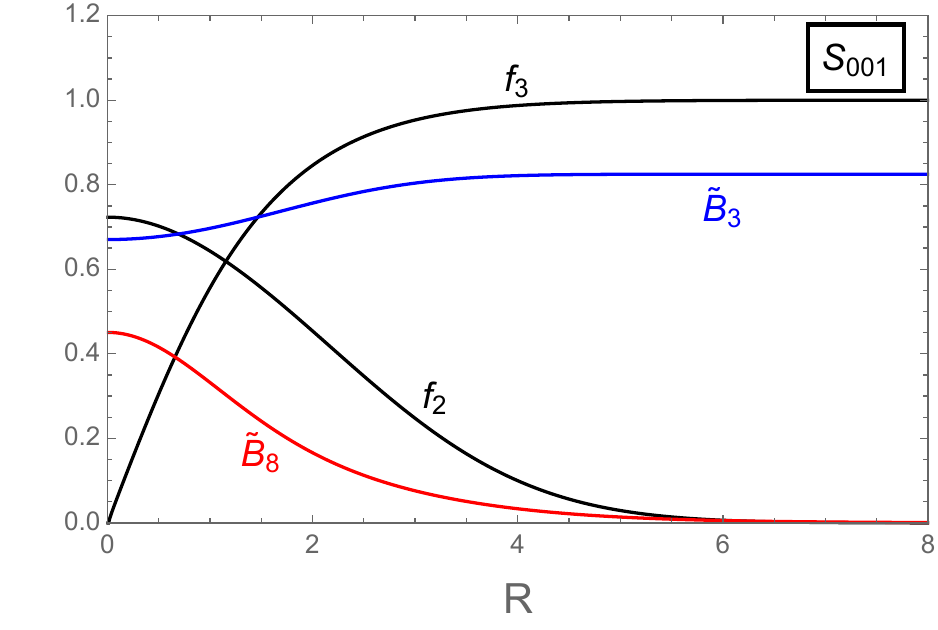}
\includegraphics[width=0.5\textwidth]{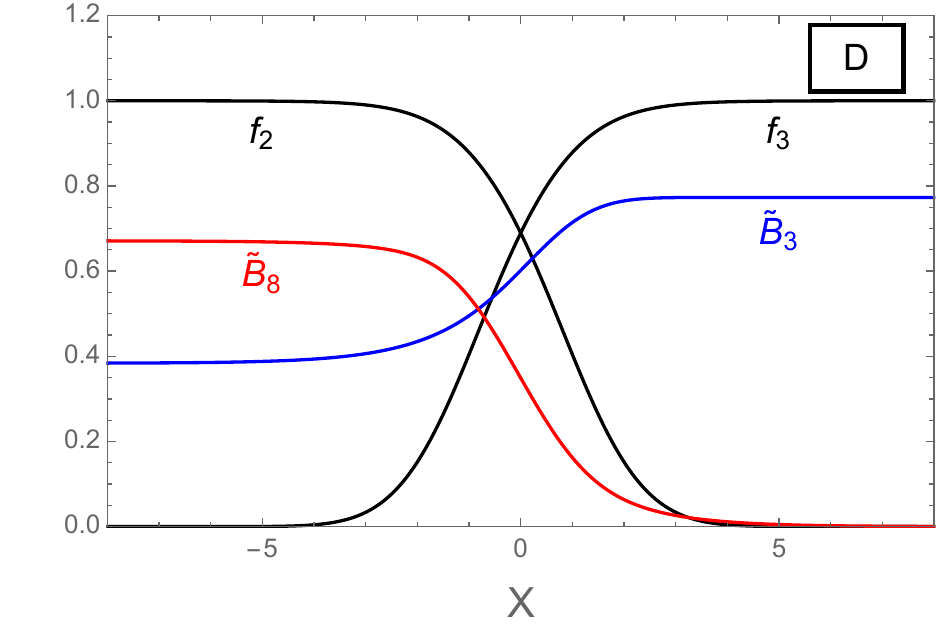}}
\caption{Profiles of the dimensionless condensates 
$f_2$, $f_3$ and the magnetic fields $\tilde{B}_3$, $\tilde{B}_8$
in units of $\mu^2/\sqrt{\lambda}$ for a 2SC flux tube (left panel, with winding number $n_3=1$) and a 2SC domain wall (right panel). The parameters for both panels are $g=3.5$, $T_c/\mu_q\simeq 0.084$,  and the profiles are plotted at their respective critical fields $H_{c1}(S_{001})=9.59\,\mu^2/\sqrt{\lambda}$ (left) and $H_{c1}(D)=8.99\,\mu^2/\sqrt{\lambda}$ (right), see also Fig.\ \ref{fig:windings}. The dimensionless radial coordinate for the flux tube is $R=r\sqrt{\lambda}\rho_{\rm 2SC}$, and the 
dimensionless cartesian coordinate $X$ for the domain wall is $X=x\sqrt{\lambda}\rho_{\rm 2SC}$. We have placed the center of the 
domain wall, where $f_2=f_3$,  at the arbitrarily chosen point $X=0$.}
\label{fig:2SC}
\end{center}
\end{figure}

\begin{figure} [t]
\begin{center}
\hbox{\includegraphics[width=0.5\textwidth]{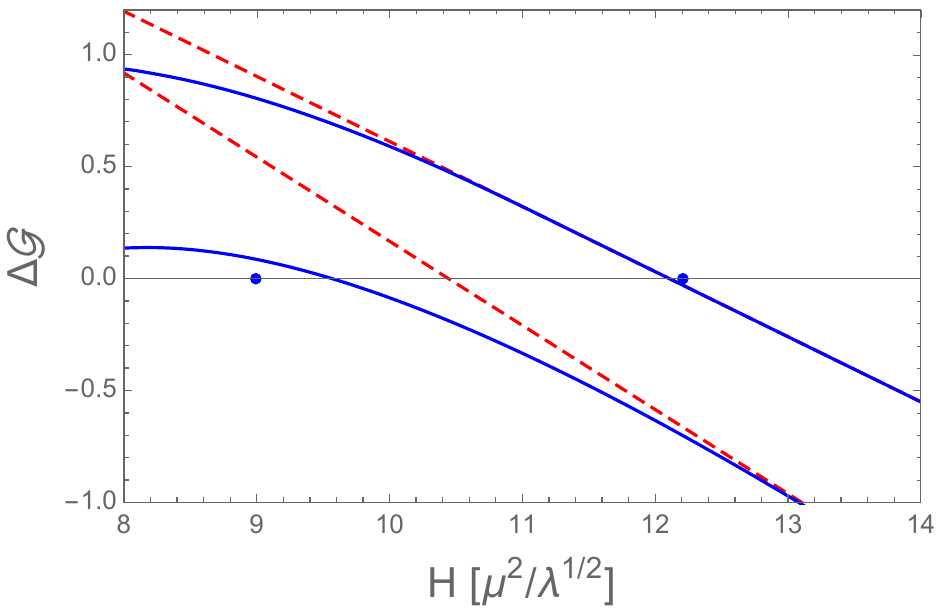}\includegraphics[width=0.5\textwidth]{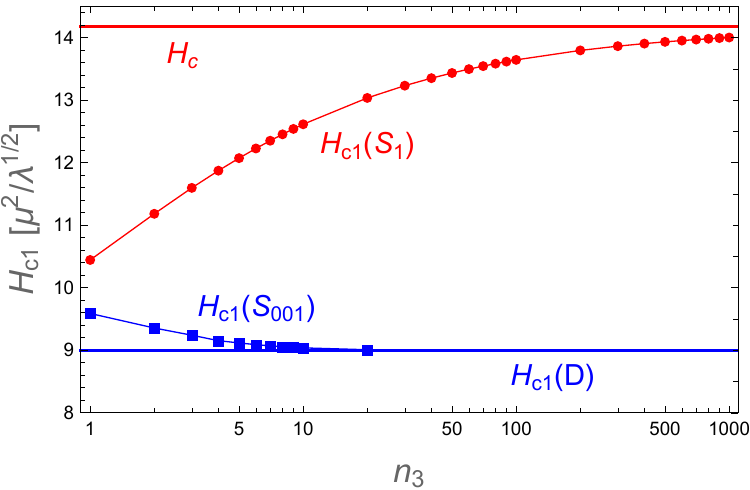}}
\caption{{\it Left panel:} Gibbs free energy difference per unit length, in units of $\rho_{\rm 2SC}^2$, between the phase with a single flux tube and the homogeneous 2SC phase ($\Delta{\cal G}=0$ defines the critical field $H_{c1}$). The solid (blue) lines
are the curves for flux tubes with an induced second condensate in the core $S_{001}$, dashed (red) lines correspond to standard flux tubes $S_1$. The lower pair of curves is computed at $T_c/\mu_q=0.084$, the upper pair at $T_c/\mu_q=0.065$, both for $g=3.5$ and winding $n_3=1$. The two dots indicate the critical fields of the domain wall. {\it Right panel:} Critical magnetic fields $H_{c1}$ for different winding numbers $n_3$ for $S_1$, $S_{001}$, and the domain wall $D$ for $T_c/\mu_q\simeq 0.084$, $g=3.5$. For large winding numbers, $H_{c1}(S_1)$ approaches $H_c$ from below, indicating ordinary type-II behavior, while $H_{c1}(S_{001})$ approaches  the critical field for the formation of domain walls from above. The thin lines connecting the data points are to guide the eye, only integer values of $n_3$ make sense.}
\label{fig:windings}
\end{center}
\end{figure}

We show the profiles for a 2SC flux tube and a 2SC domain wall in 
Fig.\ \ref{fig:2SC}. For all flux tube solutions discussed in the following, we have set the winding numbers of the components that vanish far away from the flux tube 
to zero, $n_1=n_2=0$. We have checked for some selected parameter sets that nonzero $n_1$ and/or $n_2$ give rise to less preferred configurations, which is expected because in this case $f_1$ and/or $f_2$ must vanish in the center of the tube and can only become nonzero in an intermediate radial regime.
The left panel of the figure shows a flux tube in which one additional condensate, namely $\rho_2$, is induced in the core. 
We did find parameter regions which allow for solutions where both $\rho_1$ and $\rho_2$ become nonzero in the center of the flux tube. However, we did not find any parameter region where it is energetically favorable to place a flux tube with three nonzero condensates into the homogeneous state. We shall thus ignore these configurations from now on. The configuration with two nonzero condensates, on the other hand, can become favorable over the homogeneous phase. This is shown in the left panel of Fig.\ \ref{fig:windings}, where we plot the dimensionless Gibbs free energy difference between the phase with a single flux tube and the homogeneous phase,
\be
\Delta{\cal G} \equiv \frac{G-G_{{\rm 2SC}_{\rm ud}}}{\rho_{\rm 2SC}^2 L} = \pi\left({\cal I}_\circlearrowleft -\frac{8\lambda\Xi n_3}{g^2}\right) \, ,
\ee
with $G$ from Eq.\ (\ref{G2SCflux}) and $G_{{\rm 2SC}_{\rm ud}}$ from Eq.\ (\ref{Gudhom}). The two pairs of curves show one example where the configuration with an induced condensate in the core is preferred at the point where $\Delta{\cal G}=0$ over the standard flux tube solution $S_1$, and one example where there is only a single condensate at $\Delta{\cal G}=0$. In the former case, it turns out that the system can further reduce its free energy by 
replacing $S_{001}$ with a domain wall, whose critical field $H_{c1}(D)$ is determined by solving ${\cal I}_{||}=0$ numerically for $\Xi$. This critical field is indicated in the left panel of Fig.\ \ref{fig:windings} by a dot for both cases: $H_{c1}(D)<H_{c1}(S_{001}) < H_{c1}(S_1)$ for $T_c/\mu_q=0.084$, and $H_{c1}(S_1)<H_{c1}(D)$ for $T_c/\mu_q=0.065$.
The connection between the flux tube $S_{001}$ and the 
domain wall can be understood with the help of the right panel of Fig.\ \ref{fig:windings}. Let us first explain the upper two (red) curves in this plot, which 
show the standard behavior of an ordinary type-II superconductor: the most favorable configuration is a flux tube with minimal winding number, and as we increase the winding, the 
critical field $H_{c1}$ approaches the critical field $H_c$ from below (in a type-I superconductor, it would approach it from above). This is easy to understand: as the winding is increased, the core of the flux tube becomes larger and thus the normal phase "eats up" the superconducting phase. Hence, for infinite winding, the critical field $H_{c1}$ indicates that it has now become favorable to place an infinitely large flux tube into the system, i.e., to replace the superconducting phase with the normal phase, which is nothing but the definition of $H_c$.
Similarly, the critical field for the flux tube $S_{001}$ 
approaches the critical field for the domain wall $D$: again, as we increase the winding, the phase in the core, which now approaches the 2SC$_{\rm us}$ phase 
for $n_3\to\infty$, spreads out and "eats up" the phase far away from the flux tube, which is the 2SC$_{\rm ud}$ phase. However, in contrast to the ordinary flux tube $S_1$, these two phases have the same free energy for all parameter values (in the massless limit), and there can never be a well-defined transition in the phase diagram from the homogeneous 2SC$_{\rm us}$ phase to the 
homogeneous 2SC$_{\rm ud}$ phase. Instead, we find that a stable domain wall forms, which interpolates between the two phases. 
While Figs.\ \ref{fig:2SC} and \ref{fig:windings} only show results for specific parameters, we 
study the phase diagram more systematically in the next section.

%%%%%%%%%%%%%%%%%%%%%%%%%%%%%%%%%%%%%%%%%%%%%%%%%%%%%%%%%%
\section{Phase diagrams}
\label{sec:results}
%%%%%%%%%%%%%%%%%%%%%%%%%%%%%%%%%%%%%%%%%%%%%%%%%%%%%%%%%%

\begin{figure} [t]
\begin{center}
\hbox{\includegraphics[width=0.51\textwidth]{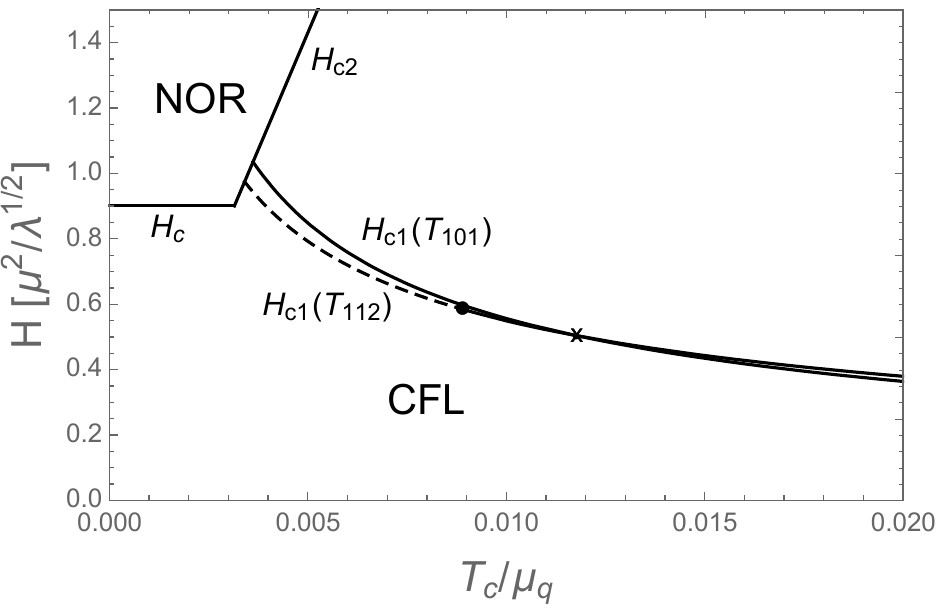}\includegraphics[width=0.49\textwidth]{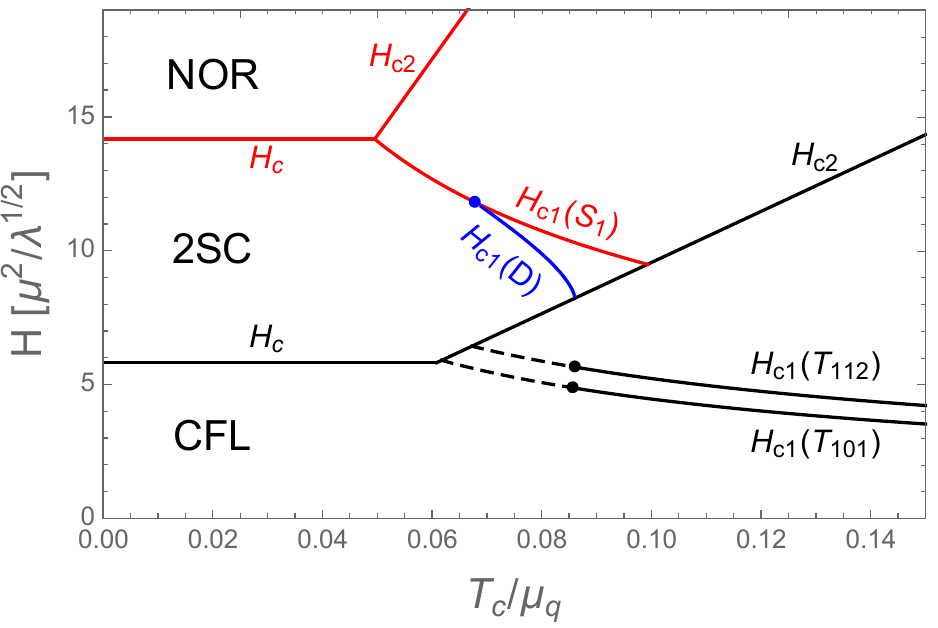}}
\caption{Critical magnetic fields for $g=0.1$ (left panel) and $g=3.5$ (right panel). For weak coupling, the 
CFL flux tube with a 2SC core $T_{101}$ is preferred over the flux tube with an unpaired core $T_{112}$  only for large 
values of $T_c/\mu_q$, while for strong coupling
this is the case for all $T_c/\mu_q$ in the type-II regime. The dots separating the 
dashed from the solid segments in $H_{c1}$ mark the transition from attractive (dashed) to repulsive (solid) long-range interaction between the flux tubes. The point marked with a cross in the left panel is  the intercept $H_{c1}(T_{101})=H_{c1}(T_{112})$. At strong coupling, where the 2SC phase appears for large magnetic fields, the 2SC domain wall $D$ is preferred over the "standard" 2SC flux tube $S_1$ for $T_c/\mu_q \gtrsim 0.07$.}
\label{fig:phases}
\end{center}
\end{figure}

Putting together the results of the previous sections, we show the phase structure of color-superconducting quark matter  in the $H$-$T_c/\mu_q$-plane in Fig.\ \ref{fig:phases}. The figure includes all three critical magnetic fields: $H_c$, 
indicating a first-order phase transition between homogeneous phases; $H_{c2}$, the lower boundary for the transition of a flux tube phase to a homogeneous phase; and $H_{c1}$, the field at which the system starts to form magnetic defects. 

As we have shown in Sec.\ \ref{sec:hom}, for small couplings $g$ the CFL phase is directly superseded by the NOR phase as we increase the magnetic field, while the 2SC phase appears as an intermediate phase for couplings 
$g>2e/\sqrt{15}$. We show one example for either case, with the larger coupling chosen such that it is realistic for the interior of neutron stars (we have not found any qualitative difference for other values of $g$ as long as
$g>2e/\sqrt{15}$). In a single-component superconductor, the critical lines $H_c$, $H_{c1}$, and $H_{c2}$ intersect in a single point, 
which marks the transition from type-I to type-II behavior, and in the type-II regime a lattice of flux tubes is expected between 
$H_{c1}$ and $H_{c2}$. This standard scenario is realized for the 2SC phase, see the intersecting (red) critical lines $H_c$, $H_{c1}(S_1)$, and $H_{c2}$ in the right panel. CFL, however, is a three-component superconductor and thus the transition region between 
type-I and type-II behavior is more complicated, see the (black) transition lines $H_c$, $H_{c1}(T_{101})$, $H_{c1}(T_{112})$,
and $H_{c2}$ in both panels which do not intersect in a single point.  Along the dashed segments of the transition lines $H_{c1}$, the long-range interaction between the flux tubes is attractive, see Sec.\ \ref{sec:inter}, and in this regime one expects a
first-order phase transition at some $H<H_{c1}$ \cite{Haber:2017kth}. For small coupling, the change from repulsive to 
attractive interaction occurs at different points for the $T_{101}$ and $T_{112}$ configurations (in the left panel, the $T_{101}$ tubes 
interact repulsively throughout the type-II regime). These points become identical for $g\gg e$, as we can see in the right panel and 
in Eq.\ (\ref{repul}). The precise structure of this type-I/type-II transition region is not the main point of this paper, and we refer the 
reader to Ref.\ \cite{Haber:2017kth} for a more detailed discussion in the context of a 
two-component superconductor; see for instance Fig.\ 5 in that reference, which suggests that flux tubes in CFL are possible also for values of $T_c/\mu_q$ smaller than indicated by the intercept of $H_c$ and $H_{c2}$. For our purpose, the main point is that for sufficiently large $T_c/\mu_q$, such that the interaction between flux tubes at long distances is repulsive,
we are in a "standard" type-II regime, and the onset of flux tubes occurs in a second-order transition.  It is this region in which 
we can compare the different critical fields $H_{c1}$ to obtain the energetically most preferred magnetic defect. 

Another complication arises in the right panel. We recall that, usually, $H_{c2}$
is the lower bound (assuming a second-order transition) for the transition of the flux tube phase to the normal-conducting phase. This 
is unproblematic in the case of the 2SC/NOR transition (upper $H_{c2}$ in the right panel). The lower $H_{c2}$ marks the transition from a CFL flux tube phase to a homogeneous 2SC phase. However, for sufficiently large $T_c/\mu_q$ we expect 2SC domain walls (or flux tubes) in the region above this $H_{c2}$. Therefore, although we have continued the curve for $H_{c2}$ into the 
region of large $T_c/\mu_q$ for completeness, the actual phase transitions (possibly between different flux tube lattices or stacks of domain walls) are beyond the scope of the present approach. 

In summary, neither panel in Fig.\ \ref{fig:phases} is a complete phase diagram and more complicated studies are necessary
to find all phase transition lines. But they serve the purpose to carefully locate the type-II regime where our main results are valid:

\begin{itemize}
\item The CFL flux tube $T_{101}$ (which has a 2SC core) has a smaller critical magnetic field $H_{c1}$ than the flux tube $T_{112}$ (which has an unpaired core), unless the strong coupling constant is very small. This is equivalent to saying that the energy 
per unit length of $T_{101}$ is smaller. Although the configuration $T_{101}$ had never been discussed before in the literature, this result is not  surprising, because the "total winding" (for instance defined by the sum of the squares of the winding numbers $n_1$, $n_2$, $n_3$) is minimized by $T_{101}$ within the constraints of a nonzero $\tilde{B}_8$-flux and a vanishing baryon circulation. 

\item The 2SC domain wall, which interpolates between the two phases 2SC$_{\rm us}$ and 2SC$_{\rm ud}$, has a lower critical field $H_{c1}$ than the standard 2SC flux tube (in which two of the three condensates are identically zero) for sufficiently large $T_c/\mu_q$. Just like the flux tube, the domain wall admits 
additional $B$-flux into the system, which is the reason it can have a lower Gibbs free energy than the homogeneous phase. 

\end{itemize} 

%%%%%%%%%%%%%%%%%%%%%%%%%%%%%%%%%%%%%%%%%%%%%%%%%%%%%%%%%%
\section{Summary and outlook}
\label{sec:summary}
%%%%%%%%%%%%%%%%%%%%%%%%%%%%%%%%%%%%%%%%%%%%%%%%%%%%%%%%%%

We have discussed magnetic defects -- flux tubes and domain walls -- in color-superconducting phases of dense quark matter,
using a Ginzburg-Landau approach. 
In a color superconductor, line defects can, in general, carry baryon circulation, magnetic flux, and color-magnetic flux.
We have focused  on the "pure" magnetic flux tubes, which have zero baryon circulation and thus are not induced by rotation. 
These flux tubes are not protected by topology, but can be stabilized by an external magnetic field. 
By solving the equations of motion numerically we have calculated the profiles of different kinds of flux tubes and their energy. As one of our main results, we have found a new type of CFL flux tube, which is most easily understood as a CFL flux tube with a 2SC core
(while the flux tube previously discussed in the literature has a core with unpaired quark matter). After carefully 
identifying the type-II regime, in which flux tubes are expected, we have shown that, for sufficiently large values of the strong coupling constant, the novel flux tube configuration has a smaller critical magnetic field 
than the flux tube with unpaired core. This result is supported by the observation that, in this strong-coupling regime,
 CFL is  superseded by 2SC as the magnetic field is increased, which makes the occurrence of CFL flux tubes with a 2SC core very 
plausible. (While, at small coupling, the CFL phase is superseded by the unpaired phase, and the flux tubes with unpaired core 
are favored.) Our new solution minimizes the total winding of the flux tube because one of the three condensates -- the one 
that survives in the 2SC phase -- has zero winding. Our second main result is the discovery of magnetic domain walls in the 2SC phase, 
which emerge from 2SC flux tubes in the limit of infinite radius. The crucial ingredient, never included in the literature before, has been to allow for induced condensates in the core of the 2SC flux tubes. We have found that one of these induced condensates grows 
until it approaches the 2SC value, giving rise to a domain wall where the profiles of the condensates interpolate between two different versions of the 2SC phase. These two versions are distinguished by the pairing 
pattern ($us$ pairing vs.\ $ud$ pairing) and have the same free energy in the limit of massless quarks, in which we have worked throughout the paper. One might argue that in this limit the 2SC phase is not relevant anyway. As we have pointed out, however, the 2SC phase can be favored over the CFL phase not only if the strange quark mass is sufficiently large, but also in the case of a large magnetic field. Therefore, the 2SC domain walls do exist in a certain regime of the phase diagram, we did not have to artificially assume the 2SC phase to be the ground state.

Nevertheless, it would be an important extension  of the present calculation to include quark masses, and, in particular, study the fate of the 2SC domain walls in this more realistic setting.  It would also be interesting to study lattices of flux tubes or stacks of domain walls rather 
than the single, isolated magnetic defects that we have studied here. We have made one step in this direction by computing the long-distance
interaction between CFL flux tubes, but a full study of inhomogeneous phases would require more involved numerical calculations. 
It is tempting to speculate about the role of the CFL flux tubes discussed here in the interior of compact stars. If a rotating neutron star has 
a CFL core, flux tubes with nonzero baryon circulation must form, because this is how a superfluid accommodates rotation. 
Since it has been shown that color neutral vortices are disfavored, these flux tubes ("semi-superfluid vortices") have nonzero 
color-magnetic fluxes. Although the total color flux of three different semi-superfluid vortices is zero, particular arrangements of 
semi-superfluid vortices with nonzero total flux are conceivable (due to the mixing of photons and gluons, this also creates 
a nonzero flux with respect to the ordinary magnetic field). However, this would imply 
alignment of rotational and magnetic axes, which contradicts observations of pulsars because the pulsating signal that we 
observe results from the misalignment of rotation and magnetic field. One solution might be the coexistence of 
semi-superfluid vortices -- aligned with the rotational axis -- and the "pure" magnetic flux tubes considered here -- aligned with the
magnetic axis. The resulting core with CFL matter would be very complicated, not unlike a nuclear matter core where vortices from superfluid neutrons  and flux tubes from superconducting protons are expected to coexist. Another  question concerns the 
boundary between quark matter and hadronic matter. It has been discussed how the 
vortices and flux tubes of nuclear matter merge with semi-superfluid vortices \cite{Cipriani:2012hr}, and it would be interesting to 
investigate this question for the non-rotational flux tubes, in particular for the flux tubes with 2SC core pointed out in this work,
 which carry an additional component of  
color-magnetic flux, on top of the flux from the rotated gluon field. {Finally, it would be interesting to further investigate the influence of the color-magnetic flux tubes and domain walls on 
the emission of gravitational waves of neutron stars. We have already mentioned the continuous emission due to color-magnetic mountains in the introduction. One could also imagine an effect 
of the color-magnetic flux tube lattice on the tidal deformability of neutron stars, which is relevant for the gravitational wave emission of neutron star mergers \cite{PhysRevLett.119.161101} (similar to a possible effect of the crust of the star \cite{Penner:2011pd} or a crystalline quark matter phase in the core \cite{Lau:2017qtz}).

\begin{acknowledgments}
We would like to thank Mark Alford, Nils Andersson, Ian Jones, David M\"uller, and Armen Sedrakian for valuable comments and discussions. 
We acknowledge support from the Austrian Science 
Fund (FWF) under project no.\ W1252, and from the {\mbox NewCompStar} network, COST Action MP1304. A.S.\ is supported by the Science \& Technology Facilities Council (STFC) in the form of an Ernest Rutherford Fellowship.
\end{acknowledgments}

\appendix

%%%%%%%%%%%%%%%%%%%%%%%%%%%%%%%%%%%%%%%%%%%%%%%%%%%%%%%%%%
\section{Flux tube interaction}
\label{app:inter}
%%%%%%%%%%%%%%%%%%%%%%%%%%%%%%%%%%%%%%%%%%%%%%%%%%%%%%%%%%

The idea behind the derivation of the long-distance flux tube interaction energy (\ref{Fint}) is to add a small correction 
to the gauge fields and the scalar fields, such that without that correction the resulting profiles are the ones for a single, 
isolated flux tube. Instead of the gauge fields themselves, one works with the following vectors, which go to zero 
as $R\to\infty$,
\begin{subequations}
\bea
\vec{Q}_3(R)&\equiv&  g\frac{a_3(\infty)-a_3(R)}{R}\vec{e}_\varphi \\[2ex]
\vec{Q}_8(R)&\equiv&  2\tilde{g}_8 \frac{\tilde{a}_8(\infty)-\tilde{a}_8(R)}{R}\vec{e}_\varphi\, .
\eea
\end{subequations}
The small perturbations are now introduced via $\vec{Q}_a=\vec{Q}_{a0}+\delta\vec{Q}_a$ ($a=3,8$) and $f_i=f_{i0}+\delta f_i$
($i=1,2,3$), and we can compute the equations of motion to zeroth and first order in the perturbations. 
Then, using these equations of motion, some tedious algebra yields the 
free energy density up to second order in the perturbations from Eq.\ (\ref{Uflux}). Writing $U_{\circlearrowleft}=U_{\circlearrowleft}^{(0)}+\delta U_{\circlearrowleft}$, we have the zeroth-order contribution
\bea
U_{\circlearrowleft}^{(0)} &=& \frac{\lambda\rho_{\rm CFL}^4}{2}\left\{\frac{\kappa_3^2}{2}(\nabla\times\vec{Q}_{30})^2+\frac{3\tilde{\kappa}_8^2}{2}(\nabla\times\vec{Q}_{80})^2
+(\nabla f_{10})^2+f_{10}^2\frac{(\vec{Q}_{30}+\vec{Q}_{80})^2}{4}+\frac{(1-f_{10}^2)^2}{2} \right. \non[2ex]
&& \left.+ (\nabla f_{20})^2+f_{20}^2\frac{(\vec{Q}_{30}-\vec{Q}_{80})^2}{4}+\frac{(1-f_{20}^2)^2}{2} + 
(\nabla f_{30})^2+f_{30}^2Q_{80}^2+\frac{(1-f_{30}^2)^2}{2} \right.\non[2ex]
&&\left. -\frac{h}{\lambda}\left[(1-f_{10}^2)(1-f_{20}^2)+(1-f_{20}^2)(1-f_{30}^2)+(1-f_{10}^2)(1-f_{30}^2)\right]\right\} \, ,
\eea
and the first- and second-order contributions, which can be written as a total derivative, 
\bea
\delta U_{\circlearrowleft} &=& \lambda\rho_{\rm CFL}^4\nabla\cdot\left\{\frac{\kappa_3^2}{2}\delta\vec{Q}_3\times\left[\nabla\times\left(\vec{Q}_{30}+\frac{\delta\vec{Q}_3}{2}\right)\right]+
\frac{3\tilde{\kappa}_8^2}{2} \delta\vec{Q}_8\times\left[\nabla\times\left(\vec{Q}_{80}+\frac{\delta\vec{Q}_8}{2}\right)\right] \right.\non[2ex]
&&\left.+\delta f_1\nabla\left(f_{10}+\frac{\delta f_1}{2}\right)+
\delta f_2\nabla\left(f_{20}+\frac{\delta f_2}{2}\right)+\delta f_3\nabla\left(f_{30}+\frac{\delta f_3}{2}\right)\right\} \, .
\eea
We can now exactly follow the steps explained in Appendix C of Ref.\ \cite{Haber:2017kth} to find 
the interaction energy for two flux tubes in a distance $R_0$ from each other,\bea \label{Fint1}
\frac{F_{\rm int}^{\circlearrowleft}}{L} &=& \int_{R_0/2}^\infty\frac{2\rho_{\rm CFL}^2 R_0dR}{\sqrt{R^2-(R_0/2)^2}}\left[-\frac{\kappa_3^2}{2}\delta Q_3\left(\frac{\delta Q_3}{R}+\delta Q_3'\right)-\frac{3\tilde{\kappa}_8^2}{2}\delta Q_8\left(\frac{\delta Q_8}{R}+\delta Q_8'\right) +\delta f_1\delta f_1'+\delta f_2\delta f_2'+\delta f_3\delta f_3'\right]\non[2ex]
&=&\int_{R_0/2}^\infty\frac{2\rho_{\rm CFL}^2 R_0dR}{\sqrt{R^2-(R_0/2)^2}}\Bigg[\frac{\kappa_3^2 g^2a_3'}{2}\frac{a_3(\infty)-a_3(R)}{R^2}+6\tilde{\kappa}_8^2\tilde{g}_8^2\tilde{a}_8'\frac{\tilde{a}_8(\infty)-a_8(R)}{R^2} \non
&&\hspace{4cm} -(1-f_1)f_1'-(1-f_2)f_2'-(1-f_3)f_3'\Bigg] \, .
\eea
where, in the second line, we have written the result in terms of the full (numerically determined) profile functions. This expression can 
be used to extrapolate the interaction energy down to smaller distances. Instead, we shall only work with the asymptotic 
result which is obtained by expressing the first line of Eq.\ (\ref{Fint1}) in terms of the asymptotic approximations to the profile functions. 
This is Eq.\ (\ref{Fint}) in the main text.

\bibliography{refs1}

\end{document}